\documentclass[11pt,a4paper]{article}
\usepackage{jheppub}
\usepackage{tikz}
\usepackage{graphicx,verbatim}
\usepackage{amsmath}
\usepackage{amsfonts}
\usepackage{amssymb}
\usepackage{psfrag}
\usepackage{accents}

\usepackage[colorinlistoftodos]{todonotes}
\usepackage{epstopdf}

\def\Lie{\mathcal{L}}

\def\scri{\mathcal{I}}
\def\scrip{\scri^{+}}
\def\scrim{\scri^{-}}
\def\scripm{\scri^{\pm}}

\def\N{\mathcal{N}}

\def\chio{\mathring\chi}
\def\phio{\mathring\phi}
\def\Y{\mathring{Y}}
\def\qo{\mathring{q}}

\def\qoub{\mathring{\underbar{q}}}
\def\qub{{\underbar{q}}}
\def\qout{\mathring{\underaccent{\tilde}{q}}}
\def\vo{\mathring{v}}

\def\ello{\mathring{\ell}}
\def\ellout{\mathring{\underaccent{\tilde}{\ell}}}

\def\Do{\mathring{D}}
\def\Dout{\mathring{\underaccent{\tilde}{D}}}
\def\Doub{\mathring{\ub{D}}}
\def\dvo{\rmd^2\mathring{V}}
\def\dvout{\rmd^2\mathring{\underaccent{\tilde}{V}}}
\def\omegao{\mathring{\omega}}
\def\epsilono{\mathring{\epsilon}}
\def\epsilonout{\mathring{\underaccent{\tilde}{\epsilon}}}

\def\Ho{\mathring{\Delta}}
\def\Houl{\mathring{\underline{\Delta}}}

\def\Yut{\underaccent{\tilde}{\mathring{Y}}}
\def\psiut{\underaccent{\tilde}{\psi}}

\def\rmd{\mathrm{d}}

\def\lub{l}
\def\lubo{l_\circ}

\def\S{\mathcal{S}}
\def\M{\mathfrak{M}}
\def\D{\mathcal{D}}

\def\H{\Delta}
\def\Hul{\underline{\Delta}}
\def\l{\ell}
\def\psiul{\underline{\psi}}
\def\alphaul{\underline{\alpha}}

\newcommand{\pb}[1]{\hbox{\lower0.5ex\hbox{${}_{\longleftarrow}$}}\kern-1.9ex{\!\!#1}}

\def\G{\mathfrak{G}}
\def\g{\mathfrak{g}}
\def\Super{\mathfrak{S}}
\def\Ver{\mathfrak{V}}

\def\Lor{\mathfrak{L}}

\def\B{\mathfrak{B}} 
\def\q{\mathfrak{q}} 
\def\Q{\mathfrak{Q}}
\def\T{\mathfrak{T}}

\def\t{\tilde}

\def\b{\bar}
\def\ub{\underbar}
\def\ul{\underline}
\def\={\hat{=}}
\def\f{\frac}

\def\be{\begin{equation}}
\def\ee{\end{equation}}
\def\ba{\begin{eqnarray}}
\def\ea{\end{eqnarray}}

\begin{document}

\title{Non-Expanding horizons:\\ Multipoles and the Symmetry Group}
 \author[a]{Abhay Ashtekar,}
\emailAdd{ashtekar.gravity@gmail.com}
\author[a]{Neev Khera,}
\emailAdd{neevkhera@psu.edu}
\affiliation[a]{Institute for Gravitation and the Cosmos \& Physics Department, Penn State, University Park, PA 16802, U.S.A.}
\author[b]{Maciej Kolanowski}
\emailAdd{maciej.kolanowski@fuw.edu.pl}
\author[b]{and Jerzy Lewandowski}
 \emailAdd{jerzy.lewandowski@fuw.edu.pl}
\affiliation[b]{Institute of Theoretical Physics, Faculty of Physics, University of Warsaw, Pasteura 5, 02-093Warsaw, Poland}

\abstract{
\noindent
It is well-known that blackhole and cosmological horizons in equilibrium situations are well-modeled by non expanding horizons (NEHs) \cite{afk,abl1,lp1}. In the first part of the paper we introduce multipole moments to characterize their geometry, removing the restriction to axisymmetric situations made in the existing literature \cite{aepv}. We then show that the symmetry group $\G$ of NEHs is a 1-dimensional extension of the BMS group $\B$. These symmetries are used in a companion paper \cite{akkl2} to define charges and fluxes on NEHs, as well as perturbed NEHs. They have physically attractive properties. Finally, it is generally not appreciated that \emph{ $\scripm$ of asymptotically flat space-times are NEHs in the conformally completed space-time}. Forthcoming papers will (i) show that $\scripm$ have a small additional structure that reduces $\G$ to the BMS group $\B$, and the BMS charges and fluxes can be recovered from the NEH framework; and, (ii) develop gravitational wave tomography for the late stage of compact binary coalescences: reading-off the dynamics of perturbed NEHs in the strong field regime (via evolution of their multipoles), from the waveform at $\scrip$.}

\maketitle

\section{Introduction}
\label{s1}

Let us begin by recalling the basic features of horizons. Event horizons (EHs) are  teleological: A 2-sphere cross-section of an EH caused by a merger or a collapse that will take place a billion years from now may well be contained the room you are sitting in, and growing! Indeed, one needs access to the full evolution of space-time into infinite future even to know if it admits an event horizon. Finally, since the event horizon is defined as the future boundary of the past of $\scri^+$, the notion is not meaningful in absence of $\scri^+$. Non-expanding horizons (NEHs) and dynamical horizons (DHs) are free of these limitations because they  are quasi-local: they refer to space-time geometry only in their immediate neighborhood (for  reviews, see, e.g., \cite{akrev,boothrev,gjrev} and for early papers with similar ideas, \cite{newman,hajicek}). NEHs represent black hole and cosmological horizons in equilibrium, while DHs describe growing (or, in the quantum theory, shrinking) black holes. In particular, you can be certain that your room does not contain a 2-sphere cross-section of an NEH or a DH. Also, since they do not refer to $\scrip$, NEHs and DHs are well-defined also in spatially compact space-times.  Numerical simulations show that, after the common horizon forms in a binary black hole merger, the initially rapid evolution of the DH slows down exponentially and is quickly well described by a perturbed NEH. A series of papers, of which this is the first, will focus on NEHs and perturbed NEHs.   

Killing horizons provide well-known examples of NEHs. However, the notion of an NEH is significantly weaker. For example, (in the asymptotically de Sitter context) the Kastor-Traschen solutions represent dynamical, multi-black hole space-times that admit NEHs which are not Killing horizons \cite{kt,kt1}. In the zero cosmological context, the Robinson-Trautman solution admits an NEH, every neighborhood of which contains radiation \cite{pc}. These simple examples are taken from explicitly known exact solutions. More generally, an initial value formulation based on double null surfaces shows that there is an infinite family of NEHs which cannot be Killing horizons because certain geometrical fields on them can be explicitly `time dependent' \cite{abl1}. Nonetheless, one can extend the zeroth and first laws of black hole mechanics to any NEH by equipping it with a suitable family of null normals \cite{afk,abl2}, thereby endowing it with the structure of a weakly isolated horizon (WIH). This can always be done on any NEH and does not entail a restriction. Thus, while the notion of NEHs is very general, they are nonetheless endowed with sufficiently rich structure for obtaining results on black hole and cosmological horizons that are of direct physical interest. For these reasons NEHs have already been studied in some detail in the literature (see, e.g., \cite{sh,afk,abl1,lp1,akrev,boothrev,gjrev} and references therein). In this paper we will extend that analysis in two directions. 

The \emph{first} extension makes use of a simple observation \cite{aasb} that has received relatively little attention so far: Each NEH $\H$ comes with a three parameter family of unit, round 2-sphere metrics $\qo_{ab}$ that are conformally related to the physical (degenerate) metric $q_{ab}$ on $\H$. Using spherical harmonics of these round metrics, we will define multipole moments that provide an invariant characterization of the NEH geometry. This construction generalizes the previous definition of multipoles that was restricted to axisymmetric NEHs \cite{aepv}. 

The \emph{Second} extension refers to symmetries. The original analysis of NEH symmetries referred to diffeomorphisms that preserve the degenerate metric $q_{ab}$, and the derivative operator $D$ on the NEH, induced by the space-time $\nabla$. The possible symmetry groups were then classified  \cite{lp1}. The pair $(q_{ab},\, D)$ varies from one space-time to another. A generic NEH admits no symmetry, and the symmetry group can be at most 5 dimensional. This is analogous to the fact that a generic metric on a 4-manifold has no isometries and the isometry group can be at most 10 dimensional. In this paper, we adopt a different viewpoint and focus on the class of \emph{all} NEHs and the \emph{universal structure} they have in common. In the analogy with 4-metrics, this  corresponds to, e.g., considering the structure shared at $\scri$ by \emph{all} asymptotically flat space-times. Because this structure is much weaker than that encoded in a \emph{given} space-time metric, the asymptotic symmetry group --the Bondi, Metzner, Sachs (BMS) group $\B$-- is infinite dimensional, much larger than the isometry group of any one metric in this class. Similarly, we will find that the group $\G$ preserving the universal structure on all NEHs is infinite dimensional, much larger than the finite dimensional symmetry groups that were previously considered. Interestingly, $\G$ is isomorphic to a one-dimensional extension of the (BMS) group $\B$. In particular, in contrast to other works \cite{aasb,cfp}, the semi-direct product between the supertranslation subgroup and the the Lorentz group is exactly the same as that of $\B$. As already remarked, in the companion paper \cite{akkl2} we will discuss charges and fluxes associated with symmetries in $\G$.

The paper is organized as follows. In Sec.~\ref{s2} we first recall the known results on NEHs and then discuss multipole moments. In Sec.~\ref{s3} we introduce the universal structure of NEHs and investigate the symmetry group $\G$ that preserves this structure. We conclude in Sec.~\ref{s4} with a summary and a discussion of applications of this framework that will be developed in forthcoming works.

As noted in the abstract, null infinity, $\scripm$, of an asymptotically flat space-time is also an NEH, albeit \emph{in the conformally completed space-time}. In a forthcoming paper, we will revisit the structures at null infinity from an NEH perspective. In essence, the framework developed in this paper can be applied to $\scripm$ after taking into account the fact that the NEH structure of $\scripm$ refers to the conformally rescaled metric rather than the physical. Any one conformal completion endows $\scripm$ with the structure of an NEH with an additional small piece of extra structure which removes the `extra' (1-dimensional) symmetry in the NEH symmetry group $\G$ and reduces it to the BMS group $\B$. Similarly, we will re-analyze the charges and fluxes at $\scrip$ using the framework developed in \cite{akkl2}.  
 
Finally, it has been suggested \cite{akrev} that there may be a correlation between the gravitational dynamics at horizons in the strong field regime, and that at $\scrip$ in the asymptotic (and therefore weak field) regime. Several analyses in the literature  \cite{schnetter_2006,Ashtekar_2013,Gupta_2018,pookkolb2020horizons,prasad2021tidal} present illustrations of such a relation. We will be able to use the close similarly between structures on NEHs and those at $\scrip$ to systematically develop gravitational wave tomography: calculating the evolution of the intrinsic geometry and angular momentum structure on the horizon, directly from the knowledge of the waveform at infinity. This analysis is likely to shed light also on the issue of correlations between the horizon structure and radiation at infinity during the long 
semi-classical phase, e.g. $ \sim 10^{76}$ years it takes for a newly formed solar mass black hole to shrink to the lunar mass.

\section{Non-expanding horizon}
\label{s2}

This section is divided into three parts. In the first, we recall salient features of the geometry of NEHs. We then turn to multipole moments that provide an invariant characterization of this geometry. In the second part we introduce necessary conceptual framework and in the third we define multipoles and discuss their salient properties.

\subsection{Geometrical structures}
\label{s2.1}

Consider a smooth space-time $(M, g_{ab})$. A 3-dimensional submanifold $\H$ of $M$ will be said to be an NEH if it satisfies the following conditions:
\begin{itemize}
\item[(i)] $\Delta$ is topologically $\mathbb{S}^2\times \mathbb{R}$ and null;
\item[(ii)] The expansion $\theta_{(\lub)}$ of every null normal $\lub^a$ to $\H$ vanishes;\, and,
\item[(iii)] $R_a{}^b\, \lub^a$ is proportional to $\lub^{\,b}$ for any null normal, where $R_{ab}$ is the Ricci tensor of $g_{ab}$. 
\end{itemize}
If these conditions are satisfied by any one null normal $\lub^a$ they are satisfied by all null normals. The first two conditions are the same as those that were introduced in the first treatments of NEHs \cite{afk,abl1,lp1}, while the third condition is a little bit weaker. The original definition assumed Einstein's equations and required that $- T_a{}^b\lub^a$ be causal and future directed if $\lub^a$ is so (which is itself a weaker version of the dominant energy condition). As shown in \cite{afk}, this requirement, together with vanishing of $\theta_{(\lub)}$ and the Raychaudhuri equation, implies that $R_{ab}\, \lub^a \lub^{\,b} =0 $ on $\Delta$. The fact that $T_a{}^b\,\lub^a$ is causal then implies condition (iii), i.e., $R_a{}^b\, \lub^a$ is proportional to $\lub^{\,b}$. To summarize, then, the dominant energy condition implies the requirement made in the previous literature \cite{afk,abl1,lp1}, and that requirement in turn implies what we are now assuming. (This slight weakening of the requirement ensures that $\scrip$ is an NEH in the conformally completed space-time.)\medskip

\begin{figure}
  \begin{center}
    \begin{minipage}{2.0in}
      \begin{center}
        \includegraphics[width=1.5in,height=2.3in,angle=0]{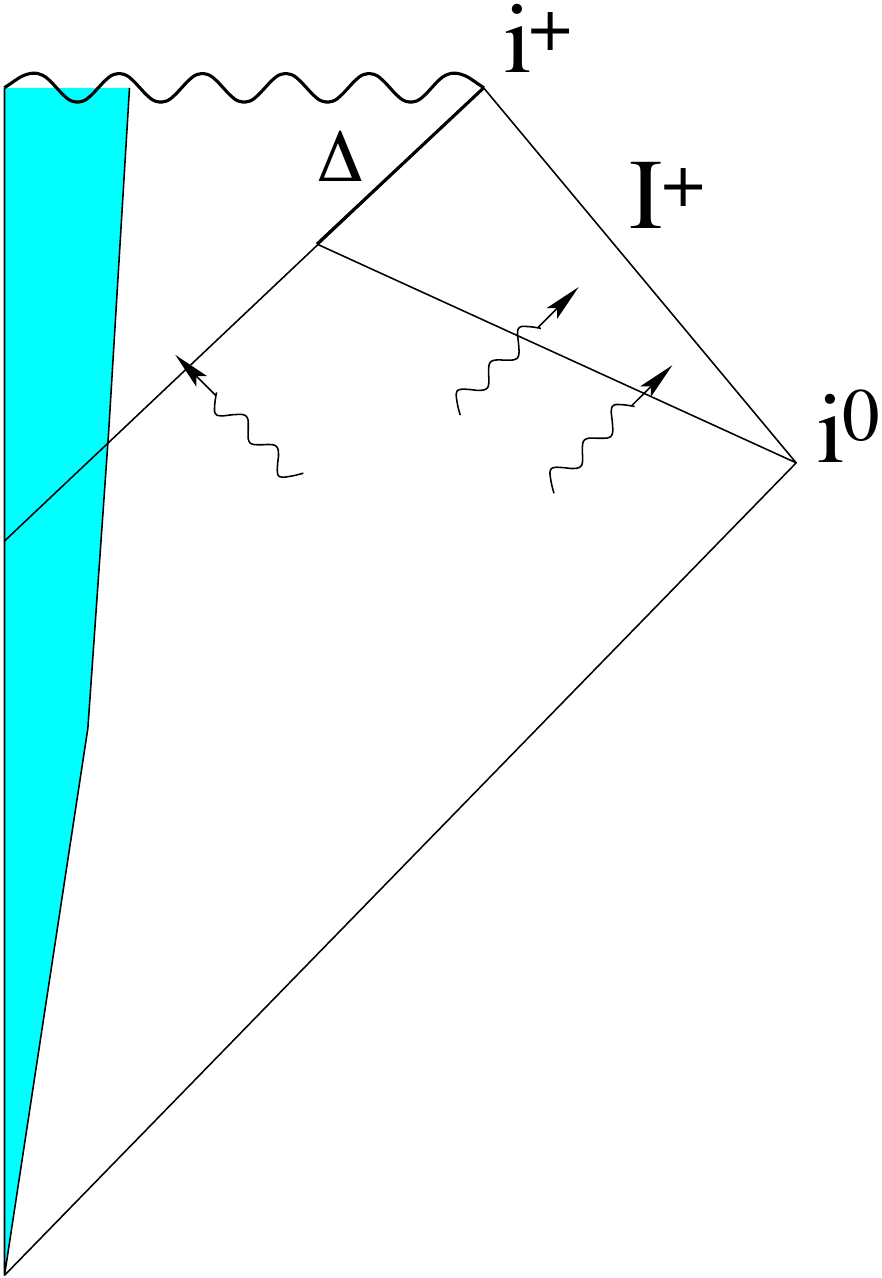}
      \end{center}
    \end{minipage}
    \hspace{.4in}
    \begin{minipage}{2.0in}
      \begin{center}\small
        \includegraphics[width=4.5cm]{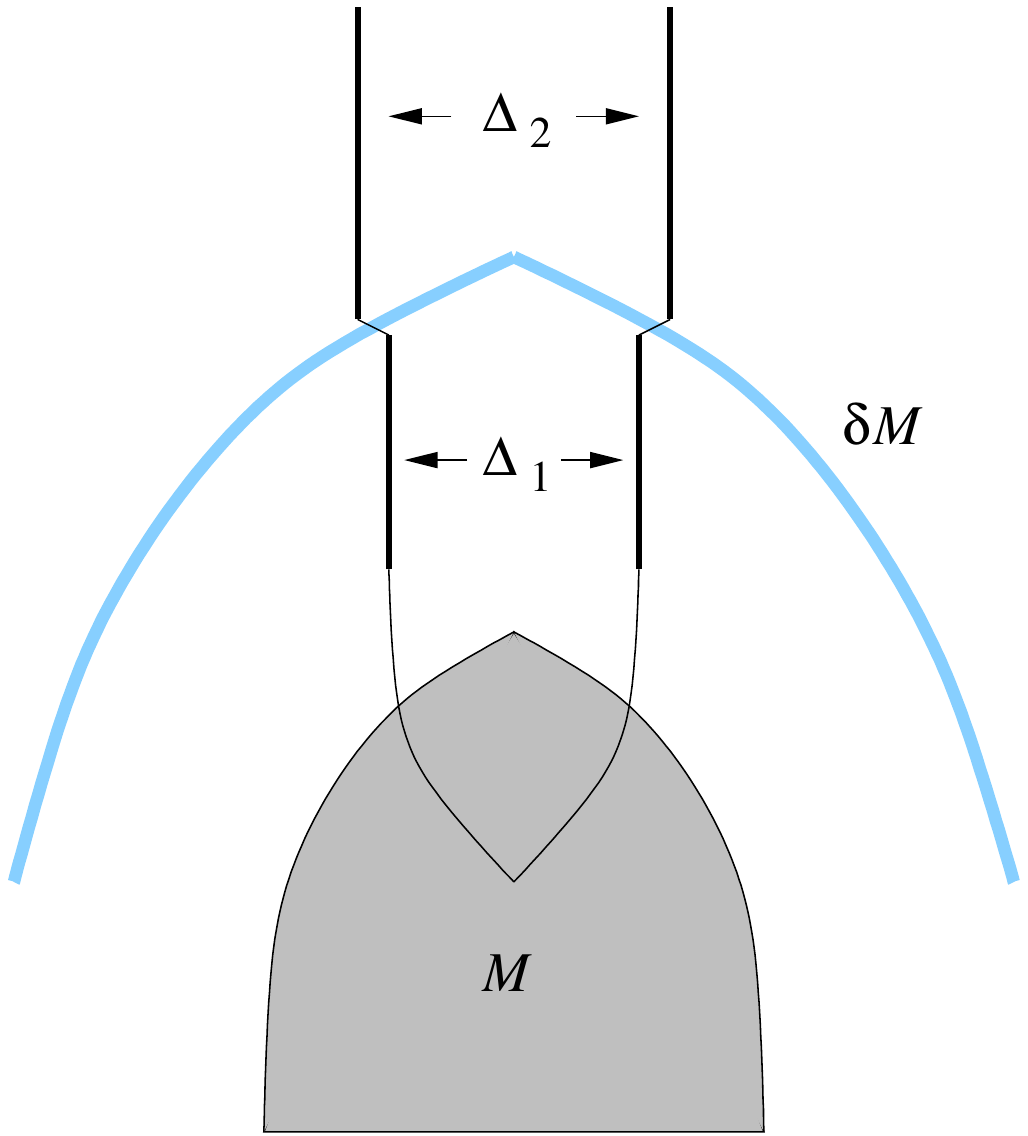}\\ 
      \end{center}
    \end{minipage}
    \caption{\footnotesize{Examples of NEHs in the asymptotically flat (i.e. $\Lambda=0$) case.\, {\emph{Left panel:}} Non-spherical gravitational collapse. Since the radiation falling into the black hole rapidly decreases at late time, the portion $\Delta$ of the event horizon is well modeled by a perturbed NEH. \, {\emph{Right panel:}} Spherical collapse of a star followed by the collapse of a spherical shell some time later. $\Delta_1$ and $\Delta_2$ are both NEHs. The event horizon lies outside $\Delta_1$ \emph{and grows} to join on to $\Delta_2$, although there is no matter or gravitational radiation falling cross it.}}\label{asymflat}
\end{center}
\end{figure}

\begin{figure}[]
  \begin{center}
  \hskip0.2cm
    \includegraphics[width=2.3in,height=2.3in,angle=0]{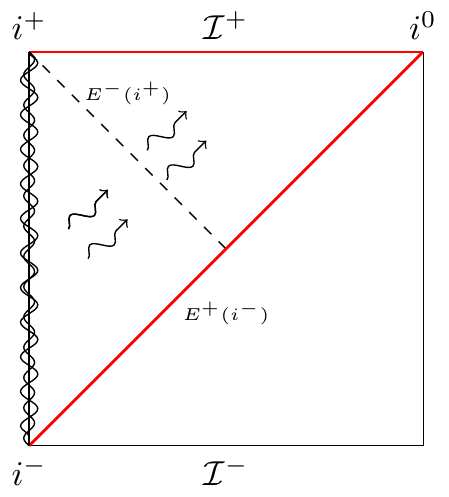}\hskip1.8cm 
    \includegraphics[width=2.4in,height=2.3in,angle=0]{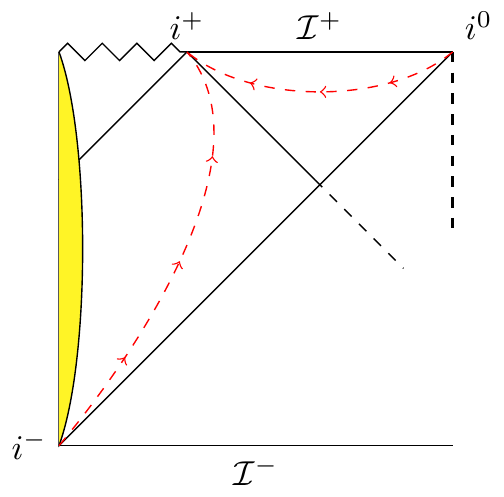}
\caption{\footnotesize Examples of NEHs in solutions with positive cosmological constant $\Lambda$. {\emph{Left panel:} Depiction of a compact binary emitting gravitational waves. The future event horizon $E^+(i^-)$ of $i^-$ is an NEH because there is no incoming radiation. If the emitted radiation is weak, the past horizon $E^-(i^+)$ of $i^+$ can be regarded as a perturbed NEH.\, {\emph{Right panel:}} Conformal diagram depicting a spherical collapse. Shaded (yellow) region corresponds to the collapsing spherical star. The  dashed (red) lines with arrows represent integral curves of the `static' Killing field. Lines at $45^\circ$ denote Killing horizons that are, in particular, NEHs. The inner NEH is the black hole horizon and the outer ones are cosmological horizons. 
  } }
\label{positivecc}
\end{center}
\end{figure}

Figures \ref{asymflat} and \ref{positivecc} provide illustrative examples of NEHs in the asymptotically flat context and in presence of a positive cosmological constant, respectively. The left panels of the two figures show compact sources that emit gravitational waves. The late time portion $\Delta$ of the event horizon in Fig.~\ref{asymflat} is an NEH if one ignores backscattering, and a perturbed NEH if one does not. In Fig.~\ref{positivecc}, the future event horizon $E^+(i^-)$ of $i^-$ is an NEH because there is no incoming radiation, while the portion of the past event horizon $E^-(i^+)$ of $i^+$ shown in the figure is a perturbed NEH, assuming that the radiation emitted by the stellar binary is sufficiently weak. The right panels of both figures depict NEHs that develop in spherical collapse. In Fig.~\ref{asymflat}, the spherical shell that collapses at late time can be replaced by a smooth profile of infalling gravitational radiation using the Chrusciel-Delay construction \cite{pced} based on gluing techniques.%
\footnote{Note incidentally, that the situation depicted in the right panel of Fig.~\ref{asymflat} has implications on what one can hope to image using ingenious techniques employed by the Event Horizon telescope. What is imaged is a snapshot of the boundary of the past domain of dependence of a dynamical horizon, which may not be a snapshot of the event horizon if the black hole of interest were to absorb significant amount of matter and radiation later on in distant future.}
The space of such solutions is infinite dimensional.

Every Killing horizon is, in particular, an NEH. But the notion of an NEH is much more general: Space-time need not admit a Killing vector even in a neighborhood of its NEH $\H$. Indeed, certain geometrical fields can be `time dependent' even on $\H$ \cite{abl1}. Nonetheless, the immediate consequences of the definition endow every NEH $\H$ with a rich structure \cite{afk,abl1,lp1}:\vskip0.1cm

1. Denote by $q_{ab}$ the pullback  $\pb{g_{ab}}$ of the space-time metric to $\Delta$. Then $q_{ab}$ is a degenerate `metric' on $\Delta$ of signature $0,+,+$ with the null normal $\lub^a$ as the degenerate direction. Further, $\mathcal{L}_{\lub}\, q_{ab} =0$ for every null normal $\lub^a$. \medskip

2. $\Delta$ is ruled by the integral curves of the null normals $\lub^{a}$. Denote by 
$\Hul$ the 2-sphere obtained by quotienting $\H$ with these curves. Properties of $q_{ab}$ imply it is the pull-back to $\H$ of a metric $\ub{q}_{ab}$ on  $\underline{\H}$. One can also pull-back to $\H$ the area 2-form $\underline{\epsilon}_{ab}$ on $\underline{\H}$, defined by the metric $\ub{q}_{ab}$. Denote it by $\epsilon_{ab}$. By definition, it satisfies $\epsilon_{ab}\, \lub^b =0$ and $\mathcal{L}_{\lub} \epsilon_{ab} =0$. \medskip

3. The Raychaudhuri equation implies that shear $\sigma_{ab}^{(\lub)}$ vanishes for every null normal $\lub^a$. It then follows that the pull-back $\pb{\nabla_{\! a}}$ to $\H$ of the space-time derivative operator $\nabla_a$ provides a well-defined (torsion-free) derivative operator $D_a$ that acts on tensor fields intrinsically defined on $\H$. \medskip

4. $D$ satisfies $D_a q_{bc} =0$ and, given any null normal $\lub^{\,b}$, we have $D_a \lub^{\,b} = \omega_a \lub^{\,b}$ for some 1-form $\omega_a$ that depends on the choice of $\lub^{\,b}$. (Thus, strictly, $\omega_a$ should carry a label $\lub$. We will omit it for notational simplicity.)  There is a rescaling freedom among the future directed null normals: $\lub^a \to \lub^{\prime \, a} = f\, \lub^a$ where $f$ is positive. Under this rescaling we have $\omega_a \to \omega^\prime_a = \omega_a + D_a \ln f$. The curl\, $2D_{[a} \omega_{b]}$\, of\, $\omega_a$\, is of course invariant under this change and, as noted below, determines ${\rm Im} \Psi_2$ at the NEH. In turn, ${\rm Im} \Psi_2$ encodes angular momentum of the given NEH. Therefore $\omega_a$ is often referred to as the rotational 1-form.
\medskip

5. The pull-back $C_{\!\pb{abcd}}$ to $\H$ of the space-time Weyl tensor $C_{abcd}$ has the property that $C_{\!\pb{abcd}}\, \lub^c =0$ on $\Delta$ for any null normal $\lub^c$. In the Newman-Penrose notation, this implies that four of the ten components of the Weyl tensor vanish on $\H$:
\be \Psi_0 := C_{abcd}\, \lub^a m^b \lub^c m^d =0\qquad {\rm and} \qquad \Psi_1 := C_{abcd}\, \lub^a m^b \lub^c n^d =0.\ee 
In the choice of the null tetrad there is some freedom even after the first vector is chosen to be a specific $\lub^a$. The conditions $\Psi_0 = 0 = \Psi_1$ at $\H$ are insensitive to this gauge freedom. Furthermore, their vanishing implies that $\Psi_2 = C_{abcd}\, \lub^a m^b \bar{m}^c n^d$ is also gauge invariant. \medskip

6. $\Psi_2$ can be expressed using geometric fields that are intrinsic to $\H$:
\be\label{psi2} {\rm Re}\, \Psi_2 =  -\f{1}{4}\,\mathcal{R}\,+\, \f{\Lambda}{6} \qquad {\rm and} \qquad({\rm Im} \Psi_2)\, \epsilon_{ab} = D_{[a} \omega_{b]}\, ,\ee 
where $\mathcal{R}$ is the pull-back to $\H$ of the scalar curvature $\underline{\mathcal{R}}$ of $\qub_{ab}$ on $\Hul$. As discussed in Sec. \ref{s2.2}, the real and imaginary parts of $\Psi_2$ determine the the geometric multipoles of $\H$ that capture the distortion in its intrinsic geometry and its angular momentum content, respectively. 
\medskip 

7. Among null normals $\lub^a$ to $\Delta$, there is a sub-family $\lubo^a$ tangential to affinely parameterized null geodesics: $\lubo^a\, \nabla_a \lubo^{\,b} = 0$ or, equivalently, $\lubo^a D_a \lubo^{\,b} = 0$, i.e., $\omega_a\lubo^a =0$ on $\H$. Thus the acceleration of these null normals vanishes on $\H$. Note that this acceleration is not an intrinsic property of the NEH geometry since its value depends on the choice of the null normal. For \emph{Killing horizons,} the acceleration of the appropriately normalized Killing vector field at $\H$ is often taken to be the surface gravity $\kappa$ of the horizon itself (and, in the quantum theory, $2\pi \hbar\kappa$ is then interpreted as the temperature seen at infinity.) The NEHs considered in this paper need not be Killing horizons. Therefore the vanishing of the acceleration has no physical significance; it just refers to a class of null normals $\lubo^{\,a}$ that can be chosen on \emph{any} NEH (including non-extremal Killing horizons that have non-zero $\kappa$).  

Any two such geodesic null normals $\lubo^{a}$ and $\lubo^{\prime\, a}$ are related by $\lubo^{\prime\, a} = f\, \lubo^{a}$, with $\mathcal{L}_{\lub} f =0$ for any null normal $\lub^a$. Now, for these normals we also have 
\be{ \mathcal{L}_{\lubo} \omega_a = 2\, \lubo^a\, D_{[a} \omega_{b]} + D_a (\omega_b\, \lubo^{\,b})\, =\, 2\, ({\rm Im}\Psi_2)\, \lubo^{\,b}\,\epsilon_{ab} =0}\, .\ee
Therefore, the 1-form $\omega_a$ corresponding to these null normals $\lubo^a$ is the pull-back to $\H$ of a 1-form $\ul{\omega}_a$ on the 2-sphere $\Hul$ of generators of $\H$. From Eq. (\ref{psi2}) we know that the curl of this 1-form is proportional to the gauge invariant scalar ${\rm Im} \Psi_2$, while its divergence on $\Hul$ changes under rescalings $\lubo^{\prime\, a} = f \,\lubo^{a}$. This `gauge freedom' can be drastically reduced by asking that $\ul{\omega}_a$ be divergence free, i.e. $\ub{q}^{ab} D_a \ul{\omega}_b =0$ on $\Hul$: The remaining rescaling freedom is just $\lubo^{\prime\, a} = c \,\lubo^{a}$ where $c$ is constant on $\H$. We will denote these distinguished null normals by $\ell^a$.

 \emph{Thus, each NEH admits a distinguished, small subclass $[\ell^a]$ of geodesic null normals}, for which $\ul{\omega}_a$ is divergence-free on $\Hul$. Since $\ul{\omega}_a$ is a divergence-free 1-form on a 2-sphere, it is co-exact: It admits a scalar potential $B$ such that 
\be \label{B} \ul{\omega}_a  =  \ul{\epsilon}_a{}^b \,\,\ub{D}_b\, B \qquad {\rm whence} \qquad {\rm Im}\Psi_2 = - \f{1}{2}\, \ub{D}^2 B.  \ee
We will refer to the scalar field $B$ as the `magnetic' potential since its Laplacian determines ${\rm Im} \Psi_2$. (Since $\mathcal{L}_\ell \Psi_2 =0$ and $\mathcal{L}_\ell B =0$,  one can think of them as function on $\H$ or on $\Hul$.) \medskip

In the rest of the paper we will make use of these geometric properties of the fields defined on any given NEH. The fact that one can only select a canonical \emph{equivalence class} of null normals $[\ell^a]$ --i.e., one cannot eliminate the ambiguity of rescaling the $\ell^a$ by a \emph{constant}-- will have interesting implications in Sec. \ref{s3}.

\subsection{Multipole moments: Conceptual Framework}
\label{s2.2}

As explained in \cite{aepv}, multipoles can be used to characterize the horizon geometry in an invariant fashion. This characterization has been used in numerical relativity to extract gauge invariant information in the horizon geometry and to compare the results of different simulations that are typically based on different choices of coordinates and gauges  (see, e.g., \cite{schnetter_2006,Ashtekar_2013,Gupta_2018,pookkolb2020horizons,prasad2021tidal}). They have also been used in entropy calculations of distorted and rotating horizons in quantum gravity (see, e.g., \cite{ashtekar_2005}). However, investigations to date assume that the horizon is axisymmetric while it is in equilibrium (or asymptotically reaches an equilibrium state that is axisymmetric \cite{Ashtekar_2013}). In this section we will provide a definition of multipoles that does not require this assumption. For simplicity we will restrict our discussion to the so-called geometric multipoles; the mass and angular momentum multipoles --with correct physical dimensions-- can be obtained from these by appropriate rescalings \cite{aepv}.

We saw in Sec. \ref{s2.1}, every NEH is equipped with two geometric fields: the degenerate metric $q_{ab}$ and an equivalence class $[\ell^a]$ of normals, where two are equivalent if and only if they differ by a \emph{constant} rescaling. $\H$ carries a natural derivative operator $D$ satisfying $D_a q_{bc} =0$ and $D_a \ell^b = \omega_a \ell^b$ for any $\ell^a \in [\ell^a]$, (where $\omega_a$ is independent of the choice of $\ell^a \in [\ell^a]$). Curvature of $q_{ab}$ and the 1-form $\omega_a$ determine $\Psi_2$ via Eq. (\ref{psi2}). Geometric multipole moments are defined using $\Psi_2$. There are two sets: the `shape' multipoles that capture the distortions of the NEH geometry are extracted from  ${\rm Re} \Psi_2 =- \f{1}{4} \mathcal{R} + \f{\Lambda}{6}$, the scalar curvature of $q_{ab}$, and the `current' multipoles, that encapsulate the rotational structure of the NEH, from ${\rm Im} \Psi_2 = \f{1}{2} \epsilon^{ab} D_a \omega_b$.%
\footnote{The inverse $\epsilon^{ab}$ of $\epsilon_{ab}$ is ambiguous up to addition of terms of the type $t^{(a} \ello^{\,b)}$ where $t^a$ is any vector field tangential to $\H$. But because of properties of $\omega_a$, the expression ${\rm Im} \Psi_2$ in terms of $\omega_a$ in insensitive to this ambiguity.}
These numbers, extracted from $\Psi_2$, provide an invariant characterization of the horizon geometry \cite{aepv}. However, to obtain them one needs a set of spherical harmonics that can be defined using the horizon geometry in an invariant fashion, i.e. using just the fields available on any given $\H$. The strategy adapted in the previous works was to restrict oneself to situations in which $q_{ab}$ is axisymmetric and its rotational Killing field Lie drags the 1-form $\omega_a$ on $\H$. Using the orbits of the Killing field and its norm, one can then construct a canonical unit, round 2-sphere metric which shares with $q_{ab}$ the rotational Killing field \cite{aepv}. The spherical harmonics $Y_{l,m}$ of the round metric were then used to extract from $\mathcal{R}$ the geometric `shape' multipoles and from $\omega_a$ the geometric current multipoles. Because of the axisymmetry assumption, only the $m=0$ moments were non-zero.

We will now provide another avenue that does not require $q_{ab}$ and $\omega_a$ to be axisymmetric. We begin by recalling that because of its 2-sphere topology, the space $\Hul$ of generators of $\H$ admits precisely a 3-parameter family of unit round metrics $\qoub_{ab}$ each member of which is conformally related to the given 
$\qub_{ab}$. Thus, 
\ba \label{qoub} \qoub_{ab} = \psiul^2 \qub_{ab} \quad &&{\hbox{\rm where $\psiul$ satisfies}} \quad 2\, (\ub{D}^2 \ln \psiul + \psiul^2) = \underline{\mathcal{R}}\nonumber\\ 
\quad &&{\hbox{or, equivalently}}\quad\,\, 2(\mathring{\ub{D}}^2 \ln \psiul  +1) = \psiul^{-2}\, \ul{\mathcal{R}}\, . 
\ea
Here we have used the conformal transformation property of scalar curvature, and the fact that the scalar curvature of a unit 2-sphere metric is $2$. The last equation admits a 3-parameter family of solutions for any $\qub_{ab}$ on $\Hul$. It follows immediately from (\ref{qoub}) that any two of these round metrics are themselves conformally related:
\be \label{conf} \qoub^\prime_{ab} = \alphaul^2 \qoub_{ab} \qquad {\hbox{\rm where\, $\alphaul$\, satisfies}} \qquad  \,({\Doub}^2 \ln \alphaul + 1) = \,\alphaul^{-2}\,  , \ee
and where ${\Doub}$ is the derivative operator compatible with $\qoub_{ab}$. Solutions to Eq. (\ref{conf}) can be written down explicitly:
\ba \label{alpha}
\alphaul^{-1} &=& \alpha_0 + \sum_{i=1}^3 \, \alpha_i\, \hat{r}^i\, , \quad{\hbox{\rm for real constants $\alpha_0$ and $\alpha_i$,\,\,  with}} \nonumber\\
 \hat{r}^i &=& (\sin\theta\cos\phi,\, \sin\theta\sin\phi,\, \cos\theta)\qquad {\rm and}\qquad -\alpha_0^2 + \sum_{i=1}^3 (\alpha_i)^2 = -1, \ea   
where $(\theta, \phi)$ are spherical polar coordinates adapted to the given round metric $\qoub_{ab}$. The eigenvalue problem for the Laplacian 
of $\qoub_{ab}$ provides us with the spherical harmonics $\Y_{l,m}$ to define multipole moments. (Note that $\alpha^{-1}$ is a (appropriately normalized) linear combination of the first four spherical harmonics of $\qoub_{ab}$.)

By pulling-back these unit round 2-sphere metrics from $\Hul$, we acquire a 3-parameter family of metrics $\qo_{ab} = \psi^2 q_{ab}$ on $\H$, where $\psi$ is the pull-back to $\H$ of $\psiul$ and hence satisfies  $\mathcal{L}_{\ell} \psi = 0$ on $\H$ (in addition to (\ref{qoub}) on $\Hul$). Any two of these metrics are conformally related via (\ref{conf}) where the function $\alpha$ on $\Delta$, being the pull-back of $\alphaul$, also satisfies $\mathcal{L}_{\ell} \alpha = 0$. Note that this construction implies that \emph{every NEH is equipped with a canonical 3-parameter family of round metrics $\qo_{ab}$}. Although the conformal factor $\psi$ relating the physical $q_{ab}$ with any one $\qo_{ab}$ varies from one NEH to another since it depends on the physical metric $q_{ab}$, the \emph{relative} conformal factors $\alpha$ relating the $\qo_{ab}$ \emph{are universal}: they make no reference to $q_{ab}$ (or any other structure) that varies from on NEH to another.  

Next, let us return to the observation that any given NEH is equipped with a pair $(q_{ab},\, [\ell^a])$. From  $q_{ab}$ we extracted a set of 3 parameter family of metrics $\qo_{ab} = \psi^2 q_{ab}$. Given any spin-dyad $m^a, \bar{m}^a$ for $q_{ab}$\, (i.e. vector fields satisfying $q_{ab} m^a m^b=0$ and $q_{ab} m^a \bar{m}^b = 1$)\, $\mathring{m}^a = \psi^{-1} m^a$ and $\mathring{\bar{m}}^a = \psi^{-1} \bar{m}^a$ is a spin-dyad for the metric $\qo_{ab}$. Thus when the metric is rescaled by a factor $\psi^2$, it is natural to rescale the contravariant vector fields by $\psi^{-1}$. With this motivation, let us consider the 3-parameter family of pairs, $(\qo_{ab},\, [\ello^a])$, with $\qo_{ab} =\psi^2 q_{ab}$,\, and $[\ello^a] = [\psi^{-1} \ell^a]$, related to the given $(q_{ab}, [\ell^a])$ by a conformal rescaling. These pairs exist on every NEH and the conformal rescaling (by powers of $\alpha$) relating any two members $(\qo_{ab},\, [\ello^a])$ and  $(\qo_{ab}^\prime,\, [\ello^{\prime\,a}])$  of this family is universal, making no reference to the specific NEH considered. 

The rescaling of $\ello^a$ has an added attractive feature in that, as we now show, it enables one to pull together the scalar curvature $\mathcal{R}$ and ${\rm Im} \Psi_2$ --the seeds of multipole moments-- in a single unified framework. Fix any one $\ello^a$ and consider the 1-form $\omegao_a$ it defines:
\be \label{omegao} D_a \ello^b = \omegao_a\ello^b \qquad {\rm with} \qquad  \omegao_a = -D_a \ln \psi + \omega_a \, \equiv\, - \Do_a E + \epsilono_a{}^b \Do_b B \ee
where we have set $E = \ln \psi$ and used (\ref{B}) to express $\omega_a$ in terms of its potential $B$.

Let us now recall from (\ref{qoub}) that the scalar curvature $\mathcal{R}$ can be expressed in terms of $\psi$ so that $E$ is a potential for $\mathcal{R}$ and hence for ${\rm Re} \Psi_2$:
\footnote{\label{fn2}For notational simplicity, here and what follows we drop the  distinction between fields on $\Hul$ and their pull-backs to $\H$. Thus, for example, $\Do^2 E$ and $\Do^2 B$ are pull-backs to $\H$ of $\mathring{\underbar{D}}^2 \ub{E}$ and $\mathring{\underbar{D}}^2 \ub{B}$ on $\Hul$.}
\be \label{R} \mathcal{R}\,\epsilon_{ab} = 2 \,(1+ \Do^2 E)\, \epsilono_{ab}\, . \ee  
Thus Eq. (\ref{omegao}) implies that the 1-form $\omegao_a$ determined by $\ello^a = \psi^{-1}\, \ell^a$ has the pleasing property that its divergence determines ${\rm Re} \Psi_2$ and its curl determines ${\rm Im} \Psi_2$. Consequently, we have:
\be \label{potentials} -\,\big(\Psi_2 -\f{\Lambda}{6}\big)\, \epsilon_{ab} = \f{1}{2} \big((1+ \Do^2 E) \, +\, i \Do^2 B \big)\, \epsilono_{ab}\, . \ee
To obtain multipoles we have to integrate this 2-form against spherical harmonics of $\qo_{ab}$ over any 2-sphere cross-section of $\H$.

Let us return to $\omegao_a$. It has two potentials: the scalar $E$ and the pseudo-scalar $B$, whose Laplacians determine the real and imaginary parts of $\Psi_2$. The factor of $1$ on the right side of (\ref{potentials}) that appears to spoil the complete symmetry between $E$ and $B$ arises only because of a topological reason. The integral of the left side of (\ref{R}) over a 2-sphere is $8\pi$\,\, --the Gauss invariant--\,\, while the integral of $\mathring{\underbar{D}}^2 E$ vanishes identically. For ${\rm Im} \Psi_2$ there is no topological obstruction because ${\rm Im} \Psi_2\, \epsilon_{ab}$ is exact. Put differently, on any NEH, ${\rm Re} \Psi_2$ has a `constant' (or $Y_{00}$) part determined by the 2-sphere topology, while the `constant part' of  ${\rm Im} \Psi_2$ vanishes identically. If we were to remove the `constant part' from ${\rm Re} \Psi_2$, the symmetry would have been exact.  (Alternatively, if we had a NUT charge, giving rise to a non-zero constant part in ${\rm Im} \Psi_2$, the symmetry would have been restored.)

To summarize, the conformal transformation $\qo_{ab} = \psi^2 q_{ab}$ leads to a natural rescaling $\ello^a = \psi^{-1}\ell^a$ because of the contravariant nature of these vector fields. This rescaling results in an unanticipated and pleasing property that the scalar curvature $\mathcal{R}$ constructed from the metric $q_{ab}$ and ${\rm Im}\,\Psi_2$ constructed from the 1-form $\omega_a$ associated with $[\ell^a]$ are brought together and encoded in the scalar potentials of the 1-form $\omegao_a$ associated with $[\ello^a]$. This synergy holds for every pair $(\qo_{ab},\, [\ello^a])$ in the 3-parameter family. The spherical harmonics of the round metrics $\qo_{ab}$ provide the necessary (invariantly defined) `weights' and the 1-form $\omegao_a$ (defined through $D_a \ello^{\,b} = \omegao_a \ell^b$) provides us with the physical field $\Psi_2$ that serve as the seeds of multipoles.

\subsection{Multipole Moments: Definition and Properties}
\label{s2.3}

This discussion is divided into three parts. In the first, we introduce the definition of multipoles --a set of numbers extracted from the NEH geometry. In the second, we show that this geometry in turn is completely characterized by this set of numbers; there is a step by step procedure to recover it from its multipoles. In the third put this discussion in a broader setting by comparing our multipoles with those available in the literature.

\subsubsection{Definition} 
\label{s2.3.1}

Fix an NEH $\H$ and any one pair $(\qo_{ab}, [\ello^a])$ from the naturally available 3-parameter family. Then we define the geometric shape and current multipoles, $I_{\ell,m}$ and $L_{\ell, m}$ via:
\begin{align}
    \begin{split}
        I_{\ell, m} &= - \oint \left( {\rm Re} \Psi_2 - \frac{\Lambda}{6} \right) \mathring{Y}_{\ell,m} d^2V = \frac{1}{2}\oint \left(1+\mathring{D}^2 E \right) \mathring{Y}_{\ell,m}d^2 \mathring{V} \\
                L_{\ell, m} &= - \oint {\rm Im} \Psi_2 \mathring{Y}_{\ell,m} d^2V = \frac{1}{2}\oint \left(\mathring{D}^2 B \right) \mathring{Y}_{\ell,m}d^2 \mathring{V}
    \end{split}
\end{align}
where the integral is taken over any 2-sphere cross-section of $\H$, \, $\rmd^2 V$ and $\dvo$ are the area elements defined by $\epsilon_{ab}$ and $\epsilono_{ab}$, and $\Y_{\ell, m}$ are the spherical harmonics defined by the unit, round metric $\qo_{ab}$ (constituting the standard orthonormal basis). 

As in the previous definitions of geometric multipoles, the monopoles are universal; they do not depend on the specific NEH considered. We have:
\be \label{monopoles} I_{0,0} = \sqrt{\pi} \qquad {\rm and}\qquad L_{0,0} = 0 \, . \ee
These properties follow from the $\mathbb{S}^2$ topology of $\Hul$ and smoothness of fields. Non-trivial information in the geometry of a specific NEH considered is encoded in all the higher multipoles with $\ell \ge 1$. For these, we have
\ba
\label{multipoles} I_{\ell,m} + i L_{\ell, m} &=& \f{1}{2}\,\oint \big[\Do^2 (E+iB)\big]\, \Y_{\ell, m}\, \dvo\nonumber\\
 &=& \f{\ell(\ell+1)}{2}\, \oint (E+iB)\, \Y_{\ell, m}\, \dvo\,\, \ea 
Note that the 2-sphere integrals in (\ref{multipoles}) refer to the area element $\dvo$ of $\qo_{ab}$ and the $\Y_{\ell, m}$'s constitute a complete, orthonormal eigenbasis of $\Do^2$ on the space of square integrable functions on the cross section. Hence $I_{\ell,m} + i L_{\ell, m}$ are independent and together they enable one to reconstruct $E+iB$ as usual:
\be \label{EB} E+iB = \sum_{\ell =1}^\infty\, \sum_{m=-\ell}^{\ell}\, \f{2}{\ell(\ell+1)} \,\big(I_{\ell, m} + i L_{\ell, m}\big)\, \Y_{\ell,m}^\star\, . \ee 

All these steps can be carried out for each choice of the naturally available unit, round, 2-sphere metrics $\qo_{ab}$. At first it may seem surprising that there is a 3-parameter family of NEH multipoles rather than a unique set as in Newtonian gravity. But note that multipoles are only meant to provide an invariant characterization of physical fields and a priori there can be many such characterizations. An early example is provided by the Geroch multipoles in static space-times \cite{geroch} and the Hansen multipoles in stationary space-times \cite{hansen}. Because the Hansen's framework does not reduce to Geroch' in static space-times, we have two separate sets of multipoles in the static case, but each suffices to determine the solution to Einstein's equations uniquely in a neighborhood of infinity (up to a diffeomorphism). In our case the situation is even simpler: multipoles defined by different $\qo_{ab}$ just correspond to expansions in different basis functions.%
\footnote{Because the conformal factors $\alpha$ relating any two unit, round metrics $\qo_{ab}$ satisfy (\ref{alpha}), one can set up a 1-1 correspondence between the 3-parameter family of $\qo_{ab}$'s and the hyperboloid of unit time-like vectors in Minkowski space. Multipoles defined by any $\qo_{ab}$ can be identified with the components of symmetric trace-free tensors tangential to the hyperboloid at that $\qo_{ab}$. The Lorentz group has a natural action on the hyperboloid and the multipole tensors transform covariantly under this action. See Remarks 1 and 2 in the Appendix).}

However, in practice it is cumbersome to deal with the 3-parameter family of multipoles and much easier to just fix this `gauge freedom' by choosing a \emph{canonical} $\qo_{ab}$. In particular, this gauge fixing would simplify comparisons between NEHs that result from distinct numerical simulations. In each simulation, one would have a well-defined, coordinate and frame independent prescription to fix the $\qo_{ab}$ and then compute multipoles.  Simulations provide realizations of the same NEH geometry --even when they use different coordinates and frames--  if and only if the multipoles agree. While there is no canonical choice of gauge-fixing, we will introduce one that is conceptually simple: Given any physical metric $q_{ab}$ on a NEH, \emph{there exists a unique unit, round, 2-sphere metric} $\qout{}_{ab} = \psiut^2 q_{ab}$ for which the `area dipole moment' vanishes, i.e., such that the $\ell=1$ components of the area element vanish:
\be \label{cm} \oint \Yut_{1,m}\, \rmd^2 V\,\, =\,\, \oint \Yut_{1, m}\, \psiut^{-2} \, \dvout =0 \, ,\ee
where, again, the integral can be carried over any 2-sphere cross-section of $\H$. The proof of this assertion is given in the Appendix. Here we only note that the 3-dimensional freedom in the choice of $\qo_{ab}$ is fixed by imposing three conditions above, one for each $m$. If one regards the positive quantity $\psiut^{-2}$ as a fictitious `mass density', the condition can be interpreted as going to the corresponding `center of mass frame'. For definiteness, from now on by  \emph{``NEH multipoles"} we will mean multipoles of Eq. (\ref{multipoles}) defined using this choice of $\qout{}_{ab}$ on the given $\H$.

\subsubsection{Reconstruction}
\label{s2.3.2}

Let us begin with the issue of reconstructing the horizon geometry from the multipoles. Suppose we are given an NEH $\H$, the associated pair $q_{ab}$ and $[\ell^a]$, and the action of $D$ on it: $D_a q_{bc} =0$ and $D_a \ell^b = \omega_a \ell^b$.  We will refer to the triplet $\big(q_{ab}, \, [\ell^a],\, \omega_a \big)$ as \emph{the NEH geometry}. It satisfies the following conditions: (i) $q_{ab}$ and $\omega_a$ are transverse to $\ell^a$, and (ii) they are Lie-dragged by $\ell^a$. Furthermore, from the scalar curvature $\mathcal{R}$ of $q_{ab}$ and curl of $\omega_a$, we can construct the 2-form \, $-\left(\Psi_2 - \frac{\Lambda}{6} \right)\, \epsilon_{ab} = \f{1}{4}\,\mathcal{R}\, \epsilon_{ab}\, +\, D_{[a} \omega_{b]}$. Finally, we can single out the canonical round metric $\qout{}_{ab}$ using (\ref{cm}) and use its spherical harmonics $\Yut_{\ell, m}$ to calculate the multipole moments $(I_{\ell,m},\, L_{\ell,m})$ of the given NEH via Eqs. (\ref{monopoles}) and (\ref{multipoles}). 

Now suppose someone hands us only these multipoles. Then, just from this set of numbers, can we reconstruct the NEH geometry? As we now show, the answer is in the affirmative. 

Let us begin with a manifold $\Ho$ --unrelated to any space-time-- which is topologically $\mathbb{S}^2\times \mathbb{R}$ and introduce on it a ruling by integral curves of a vector field $\ello^a$. On the 2-sphere of these integral curves, fix a unit round metric and pull it back to a degenerate metric $\qo_{ab}$ on $\Ho$; thus $\qo_{ab} \ello^b =0$ and $\mathcal{L}_{\ello} \qo_{ab} =0$. Let $[\ello^a]$ denote the equivalence class of vector fields related to $\ello^a$ by a constant rescaling. Using the given set of multipoles, introduce on $\Ho$ a complex function $E+iB$ using the spherical harmonics of $\qo_{ab}$ via Eq. (\ref{EB}).
 
Next, let us set $\psi = e^{E}$. Since $\qo_{ab}$ was any unit, round, 2-sphere metric, generically the triplet $(\Y_{\ell, m}, \psi, \dvo)$ will not satisfy the condition (\ref{cm}) requiring that the `area dipole moment' should vanish. However, by construction (and the existence and uniqueness result of the Appendix), it follows that there exists a unique conformal rescaling $\qo_{ab} \to {\qout}{}_{ab} = \alpha^2 \qo_{ab}$ with $\alpha$ given by (\ref{alpha}) that satisfies (\ref{cm}):
\be \oint \Yut_{1,m}\, \psiut^{-2}\, \dvout = 0 \qquad {\rm with} \qquad \psiut = \alpha\, \psi\, . \ee
Finally, using this $\psiut$ and $B$ (which is insensitive to the choice of the unit, round metric $\qo_{ab}$), we can introduce the desired triplet $(q_{ab},\, [\ell^a],\, \omega_a)$ on $\Ho$:
\be q_{ab} = \psiut^{-2}\, {\qout}{}_{ab}\,; \qquad [\ell^a] = [\psiut\, \ellout^a]\,; \qquad {\rm and} \qquad \omega_a =  {\epsilonout}^b{}_a\, \Do_b B \, .\ee
(Note that because $\qo_{ab}$,\,\,$\qout{}_{ab}$ and $q_{ab}$ are all conformally related, ${\epsilonout}^b{}_a\, \Dout_b B = {\epsilono}^b{}_a\, \Do_b B = {\epsilon}^b{}_a\, D_b B$.) Since $\Lie_{\ello}\, \psiut =0$, it follows that $q_{ab}$ and $\omega_a$ are lifts to $\Ho$ of $\qub_{ab}$ and $\ul{\omega}_{a}$ on $\Houl$. We now have the NEH $\H$ we began with, as well as the abstract $\Ho$ we introduced, and the corresponding 2-spheres $\Houl$ and $\Hul$. Our construction implies that there is a diffeomorphism $\ul\Lambda$ from the 2-sphere $\Houl$ to the 2-sphere $\Hul$\, --defined by the $(\theta,\phi)$ charts of $\qout{}_{ab}$ on $\Hul$ and $\Houl$-- \, that sends $\ul{\mathcal{R}}$ and ${\rm Im} \ul\Psi_2$ on $\Houl$ to those on $\Hul$. (These charts were used in defining the multipoles on $\H$, and to reconstruct $E,\, B$ on $\Ho$.)

To conclude the reconstruction, note first that it is easy to extend $\ul\Lambda$ to a diffeomorphism $\Lambda$ from $\Ho$ to the given $\H$ such that it also maps the equivalence class $[\ello^a]$ on $\Ho$ to that on $\H$. This diffeomorphism sends the reconstructed triplet $(q_{ab},\, [\ell^a],\, \omega_a)$ on $\Ho$ to the triplet we started with, on the given NEH $\H$ we began with.

\subsubsection{Relation to other notions of multipoles}
\label{s2.3.3}

The multipoles defined here are to be regarded as the `source multipoles' --where the `source' is the NEH-- rather than the field multipoles defined at infinity in stationary space-times \cite{geroch,hansen}. The two sets are not related in a simple manner because, due to non-linearities, the gravitational field between the horizon and infinity also contributes to multipoles. (For the Kerr horizon, these differences are discussed in \cite{aepv}.) In this respect, our approach is in the same spirit as that of \cite{aepv,Ashtekar_2013,Owen_2009} which discuss horizon multipoles.

As we already mentioned, there are two sets of multipole moments: The geometrical shape and current multipoles (that provide an invariant characterization of the NEH geometry), and the mass and angular momentum multipoles 
(that carry the traditional physical dimensions). 
As explained in \cite{aepv} one can obtain the second set from the first through an appropriate dimensionfull rescaling that uses the horizon radius and its angular momentum. One can use the same procedure to define the mass and angular momentum multipoles in our case.

The previous work most closely related to ours \cite{aepv} assumes that the pair $(q_{ab},\, \omega_a)$ is axisymmetric. For the angular momentum dipole moment, this conceptually significant restriction was removed in \cite{korzynski} using the family of round metrics $\qo_{ab}$, conformally related to the physical $q_{ab}$ on $\H$. Our work was motivated by the desire to remove the restriction for \emph{all mass and angular momentum multipoles.}

Even if one restricts the present framework to axisymmetry, there are some technical differences. While they are not significant for applications, say to numerical relativity, we list them for completeness. First, the previous work uses isolated horizons (IHs) with non-zero surface gravity, which are more restricted than the NEHs considered here: every IH is an NEH, but the converse is not true; certain geometric fields that are time \emph{in}dependent on an IH can have time dependence on an NEH (as discussed in \cite{abl1}). Second, in the reconstruction procedure of \cite{aepv}, in addition to the multipoles, one has to provide the horizon radius. This is not necessary in our case because the round metric $\qo_{ab}$ is \emph{conformally related} to the NEH metric $q_{ab}$, whence the conformal factor $\psiut$ knows about the radius. Third, one can ask whether \emph{any} given set of multipoles can lead to a horizon geometry of an NEH or an IH. It is clear that a necessary condition is that the multipoles should decay for large $\ell$ so that the appropriate sums (in our case (\ref{EB})) converge. However, in the previous work, an additional and rather awkward inequality has to be satisfied because of the assumption of axisymmetry (see Eq. (3.10) of \cite{aepv}); the present framework does not need such additional conditions. Finally, in axisymmetric space-times the multipoles defined in Sec. \ref{s2.3.1} need not reduce to those defined in \cite{aepv}. As discussed above this is not a `problem' because each set provides an invariant characterization of the horizon geometry.  Qualitatively this is similar to the status of the Geroch's moments for static space-times \cite{geroch} and the restriction of Hansen's moments \cite{hansen} from the stationary to the static case. Nonetheless, one difference is noteworthy. In the approach developed in \cite{aepv}, the mass dipole moment is guaranteed to vanish. By contrast, in our approach it does not vanish in general; instead it is the `area dipole' (\ref{cm}) that vanishes. Could we have chosen our `canonical' unit, round metric $\qout{}_{ab}$ differently and asked, instead, that the mass dipole should vanish? As discussed in the Appendix, there do exist unit, round metrics for which the mass dipole vanishes. However, in general that metric is not unique and further conditions are needed to eliminate the freedom. Positivity of $\psi^{-2}$ is used in the proof of uniqueness in the Appendix, and $-\psi^{-2}\,{\rm Re} \left(\Psi_2 - \frac{\Lambda}{6} \right)$\, --which would replace $\psi^{-2}$ if we required vanishing of mass dipole in place of `area dipole'-- \, is not positive in general. In the Kerr family for example, it is positive if and only if $a < (\sqrt{3}/2) M$ (i.e., if and only if the horizon radius $R_{\rm Hor}$ is between $2M$ and $3M/2$;\, interestingly, $3M/2$ is halfway between the Schwarzschild value $2M$ the extremal limit, $R_{\rm hor}= M$). \\

\emph{Remark:} Since the focus of previous discussions of multipoles \cite{aepv,Ashtekar_2013,Owen_2009} was on non-extremal IHs, and since these are more commonly used in the numerical relativity literature than NEHs, let us briefly discuss the newly defined multipoles in that context. A non-extremal IH  $\b\H$ is an NEH equipped with a preferred equivalence class of null normals $[\b\ell^a]$ whose acceleration is non-zero, and which satisfy $\Lie_{\b\ell}\, q_{ab} =0$ \emph{and} $[\Lie_{\b\ell},\, D_a] t^b =0$ for \emph{all} vector fields  $t^a$ that are tangential to $\b{\H}$. (Again, the equivalence class consists of null normals that are rescalings of each other by some positive constant). The triplet $(q_{ab},\, [\b\ell^a],\, D)$ constitutes the IH geometry. By contrast, while on an NEH we can always choose null normals $l^a$ that have non-zero acceleration and satisfy $\Lie_{l} q_{ab} =0$ as well as $[\Lie_{l}, D_a] l^b =  (\Lie_{l}\, \omega_a ) l^b =0$, generically there is no  $\b\ell^a$ that satisfies $[\Lie_{\b\ell},\, D_a] t^b =0$ for \emph{all} $t^a$ tangential to $\H$. Thus, there are NEHs that do not admit an IH structure. This may suggest that multipole moments of an IH would be more restricted than those of an NEH. 

However, this is not the case for the following reason. Let us restrict ourselves to multipoles for which the right side of (\ref{EB}) converges. As we saw in Sec.~\ref{s2.3.2}, given any set  $(I_{\ell,m},\, L_{\ell,m})$, one can construct an NEH geometry $(q_{ab}, [\ell^a], \omega_a)$. However, there is additional information in $D$, contained in its action $D_a n_b$ on 1-forms $n_a$ on $\H$, satisfying $n_a\l^a\not= 0$, that is not part of the NEH geometry. This extra information in $D$ may be such that there is no $\b\ell^a$ satisfying $[\Lie_{\b\ell},\, D_a] t^b =0$ for \emph{all} $t^a$. By contrast, on a non-extremal IH, Einstein's equations imply that 
$D$ is completely determined by $(q_{ab}, \, \b\omega_a)$ \cite{abl1}; there is no `extra information'. Therefore, given \emph{any} set of multipoles, the procedure given in Sec.~\ref{s2.3.2} can be trivially extended to construct a non-extremal IH geometry $(q_{ab},\, [\b\ell^a],\, D)$ with those multipoles. To summarize, while NEHs are more general than IHs, their geometry $(q_{ab},\, [\ell^a],\, \omega_a)$ contains `less information' than the IH geometry $(q_{ab},\, [\b\ell^a],\, D)$, and there is `as much gauge invariant information' in either as there is in the set of multipoles $(I_{\ell,m},\, L_{\ell,m})$. \\

Finally, in non-equilibrium situations such as black hole mergers, the NEH (or the IH) is replaced by a dynamical horizon (DH) which is space-like and naturally foliated by marginally trapped surfaces (MTS). Each MTS carries an intrinsic 2-metric $q_{ab}$ whose scalar curvature $\mathcal{R}$ carries the information about `distortions' in its intrinsic geometry, and a 1-form $\omega_a$ that carries information about its `rotational state'. Both these fields change as time evolves, i.e., one passes from one MTS to the `next'. One might envisage characterizing this evolution in an invariant way via appropriately defined multipoles. This requires the introduction of suitable basis functions --analogs of the spherical harmonics used in the equilibrium situations. There are two approaches \cite{Owen_2009, Ashtekar_2013} to select these and the relative merits are discussed in \cite{Ashtekar_2013}. The definition of NEH multipoles presented here admits an extension to DHs and it would be interesting to investigate it in detail because  the procedure to define the spherical harmonics basis is likely to be simpler than either of the existing ones. Again, these multipoles would provide an invariant way to compare the DHs that arise in distinct numerical simulations.

\section{Universal Structure and symmetries of NEHs}
\label{s3}

In this section we will discuss features shared by \emph{all} NEHs. Thus, the emphasis will shift from the geometry of individual NEHs to structures that provide a \emph{kinematical} arena to extract physics from this geometry. At null infinity, for example, the universal structure of $\scri$ and its symmetry group --the BMS group $\B$-- provide  the kinematical arena that is common to all asymptotically flat space-times. The Bondi news tensor and the asymptotic Weyl curvature are geometric fields that vary from one space-time to another. Physical results --e.g., the expression of energy-momentum charges and fluxes and their properties-- are obtained by combining the geometric fields, that vary from one space-time to another, with the universally available kinematical symmetries in $\B$ \cite{aa-yau}. The situation is completely analogous for the NEHs. In Sec.~\ref{s3.1} we will discuss the universal structure and in Sec.~\ref{s3.2} the symmetry group $\G$ that preserves it. In the companion paper \cite{akkl2} we use the symmetries and the geometrical fields that depend on specific NEHs to extract physical quantities --charges and fluxes, associated with the symmetries in $\G$.

\subsection{Universal Structure of NEHs} 
\label{s3.1}

Recall from Sec.~\ref{s2.1} that any one NEH is naturally equipped with a metric $q_{ab}$, a ruling by null normals $\lub^a$ and a derivative operator $D$. From this triplet one can obtain a number of other fields --especially $\Psi_2$-- that encode the NEH geometry. However, these fields vary from one NEH to another. In Sec.~\ref{s2.2} we found that, although it is not obvious at first, every NEH also carries additional fields --pairs $(\qo_{ab},\, [\ello^a])$ consisting of a round, unit 2-sphere metric and an equivalence classes of null normals-- that play a key role in defining multipoles. Interestingly, although their introduction on any specific NEH requires the knowledge of structures that vary from one NEH to another, \emph{relations between these fields themselves can be specified in a universal fashion}, without referring to any specific NEH.\smallskip

Therefore, we will take the universal structure to be the following. Consider a 3-manifold $\Ho$ diffeomorphic to the product $\Houl \times \mathbb{R}$ where $\Houl$ is a 2-sphere. Equip it with pairs $(\qo_{ab},\, [\ello^a])$ such that:\smallskip

(i) $[\ello^a]$ is an equivalence class of complete complete vector fields on $\Ho$ where two are equivalent if they differ by a rescaling by a positive constant. Vector fields $\ello^a$ in all equivalence classes share the same integral curves, and the quotient of $\Delta$ by these integral curves is the 2-sphere $\Houl$. Thus, $\Ho$ admits a fixed fibration.\smallskip

(ii) Each $\qo_{ab}$ is the pull-back to $\Ho$ of a unit, round 2-sphere metric $\qoub_{ab}$ on $\Houl$. Thus $\qo_{ab} \ello^b =0$ and $\mathcal{L}_{\ello} \qo_{ab} =0$ on $\Ho$.\smallskip 

(iii) Any two pairs $(\qo_{ab},\, [\ello^a])$ and $(\qo_{ab}^\prime,\, [\ello^{\prime\,a}])$ are related by
\be \qo^\prime_{ab} = \alpha^2 \qo_{ab} \qquad {\rm and} \qquad  [\ello^{\prime\,a}] =   [\alpha^{-1}\,\ello^a]\qquad {\rm with} \qquad \mathcal{L}_{\ello}\, \alpha =0.\ee

Since $\Ho$ is an abstract 3-manifold that is not embedded in any space-time, we will make a few remarks to clarify the available structure.\smallskip

(1) Note that there is no requirement that $\ello^a$ be affinely parameterized geodesic vector fields, or that they be null, because these notions refer to the 4-metric which is not universal. 

(2) However, $\Ho$ is complete in the sense that the vector fields $\ello^a$ are required to be complete vector fields. Note that if any one $\ello^a$ is complete, so are all because $\mathcal{L}_{\ello}\, \alpha =0$. A `concrete' NEH $\H$ need not be complete. $\Ho$ provides a universal home for all these concrete NEHs in the sense that they can all be embedded into $\Ho$ in such a way that the pairs $(\qo_{ab},\, [\ello^a])$ on $\H$ are mapped to those on $\Ho$. Any two embeddings are related by a horizon symmetry -- i.e., a diffeomorphism on $\Ho$ that preserves the universal structure thereon.  

(3) Because each $\qo{}_{ab}$ on $\Ho$ is the pull-back of $\qoub_{ab}$ on $\Houl$, there a derivative operator $\Do$ that acts on covariant tensor fields $t_{a_1\ldots a_n}$ on $\Ho$ that are pull-backs of tensor fields $\ub{t}_{a_1\ldots a_n}$, i.e. that satisfy $ t_{a_1,\ldots a_n} \ello^{a_1} =0,\,\, \ldots \,\, , t_{a_1,\ldots  a_n} \ello^{a_n} =0$ and $\mathcal{L}_{\ello}\, t_{a_1\ldots a_n}=0$ for any $\ello^a$ in the collection: $\Do_b\, t_{a_1\ldots a_n}$ is the pull-back to $\H$ of $\Doub{}_b\, \ub{t}_{a_1\ldots a_n}$, where $\Doub{}_a \,\,\qoub{}_{bc} =0$. 

(4) Each $\qo_{ab}$ is a degenerate metric with signature 0,+,+ on $\Ho$. A symmetric contravariant tensor field $\qo^{ab}$ will be said to be an inverse of $\qo_{ab}$ if it satisfies $\qo^{ab}\,\qo_{ac}\,\qo_{bd} = \qo_{cd}$. Thus, the inverse is ambiguous up to additions of a term of the type $t^{(a} \ello^{b)}$. 

(5) If $\mathcal{L}_{\ello}\, f =0$ then $\Do^2 f := \qo^{ab} \Do_a \Do_b f$ is well-defined, i.e., is insensitive to the ambiguity in the choice of the `inverse metric' $\qo^{ab}$. 

(6) Since each $\qoub{}_{ab}$ is a unit, round 2-sphere metric, the conformal factor $\alpha$ relating any two must satisfy ($\mathcal{L}_{\ello}\, \alpha = 0$\, and)\, $\Do^2 \ln \alpha\, +\, 1 = \,\alpha^{-2}$. There is precisely a 3-parameter family of solutions to this equation, given explicitly in  Eq.~(\ref{alpha}) in terms of the first 4 spherical harmonics of $\qo_{ab}$. 

(7) Interestingly, these considerations are parallel to those at future null infinity $\scrip$ of an asymptotically flat space-time \cite{aa-yau}. There, one specifies the universal structure also on an abstract 3-manifold --which is not embedded in any 4-dimensional space-time. The  pairs $(\qo_{ab},\, [\ello^a])$ are replaced by $(\qo_{ab}, \mathring{n}^a)$ in Bondi conformal frames. In the case of null infinity, again, the abstract $\scrip$ is required to be complete but there are solutions to Einstein's equation (such as the c-metric \cite{aatd}) in which the concrete $\scrip$ fails to be complete. The one difference between an NEH and $\scrip$ is that $\scrip$ carries vector fields $n^a$ in place of equivalence class $[\ello^a]$. As we will see, this difference leads to a 1-dimensional enlargement of the NEH symmetry group $\G$\, relative to the BMS group $\B$ at $\scrip$.

\subsection{Symmetry Group of NEHs}
\label{s3.2}

This section is divided into three parts. In the first we introduce  the symmetry Lie algebra $\g$ and the symmetry group $\G$; in the second we discuss salient properties first of $\g$ and then of $\G$; and in the third we compare $\G$ with the BMS group $\B$ and also with similar symmetry groups associated with internal boundaries that have been introduced in the literature \cite{aasb,donnay2016,hawking2017,cfp,freidel2021}.

\subsubsection{Symmetry Vector fields $\xi^a$} 
\label{s3.2.1}

The symmetry group $\G$ is the subgroup of the diffeomorphism group ${\rm Diff} (\Ho)$ that preserves the universal structure. As is often the case, it is convenient to begin with the Lie-algebra $\g$ of $\G$. It is generated by vector fields $\xi^a$ on $\Ho$ that map any given pair $(\qo_{ab},\, [\ello^a ])$ in the collection to another. Recall that any two metrics $\qoub{}_{ab}$ and $\qoub{}_{ab}^{\prime}$ in this collection are related by a conformal transformation $\qoub{}_{ab}^{\prime} = \alpha^2 \qoub_{ab}$, where $\alpha$ satisfies (\ref{alpha}). Therefore, along the 1-parameter family of diffeomorphisms generated by $\xi^a$, the given metric $\qoub{}_{ab}$ on $\Houl$ is mapped to $\qoub{}_{ab}^{\prime}(t) = \ul{\alpha}^2(t) \qoub_{ab}$, and $[\ello^a]$ is mapped to $[\alpha^{-1}(t)\ello^a]$,\, where $\ul{\alpha}(t)$ satisfies  $\Doub^2\,\ln \ul{\alpha} (t)\, +\, 1  = \ul{\alpha}^{-2}(t)$\, and\, $\alpha(t)$\, is as usual the pull-back of $\ul\alpha (t)$ to $\Ho$. Taking the derivative w.r.t. $t$,  and evaluating at $t=0$ we obtain
\be \label{main} \Lie_\xi\, \qo_{ab} = 2\phio\, \qo_{ab}\qquad {\rm and} \qquad \Lie_\xi\, \ello^a = -\big(\phio + k\big)\,\ello^a \ee
where $k$ is a constant and $\phio = \rmd \alpha(t)/ \rmd t \mid_{t=0}$ satisfies
\be \label{Y1m} \Doub^2 \phio + 2\phio =0\, , \ee 
obtained by linearization the equation satisfied by $\ul\alpha(t)$. The constant $k$ arises because the diffeomorphism has only to map\, $[\ello^a]$\, to\, $[\alpha^{-1} \ello^a]$\, rather than each $\ello^a$ to $\alpha^{-1}\, \ello^a$. Thus, if $\phio$ were to vanish, that symmetry vector field would leave each $\qo_{ab}$ invariant but can still rescale any one $\ello^a$ by a constant.

Since $\phio$ and $k$ refer to the symmetry vector field $\xi^a$, they should be denoted by $\phio_{(\xi)}$ and $k_{(\xi)}$ but for notational simplicity we will leave out the subscript when there is no possibility of confusion. The symmetry Lie algebra $\g$ consists of vector fields $\xi^a$ satisfying (\ref{main}), the Lie bracket being given by the Lie derivative of one symmetry vector field with respect to another. We will analyze the structure of this Lie algebra step by step.\smallskip

Let us first consider the \emph{`vertical'} symmetry vector fields $V^a = f \ello^a$. Then, the first of  Eqs.(\ref{main}) implies $\phio =0$ and then the second implies $\Lie_{\ello}\, f = k$, whence
\be \label{vertical} f = k \, \vo\, + \mathring{s} \qquad {\rm where} \qquad  \ello^a\partial_a\, \vo =1, \qquad {\rm and}\qquad \Lie_{\ello}\,\mathring{s}=0. \ee
(Again we have left out the subscript on $f_{(V)},\, k_{(V)}$ and $\mathring{s}_{(V)}$.) The decomposition $f= k\, \vo\, + \mathring{s}$ is not canonical but depends on the choice of the affine parameter $\vo$ of $\ello^a$. Under the transformation of the affine parameter\, $\vo \to \mathring{\bar{v}} = \vo + a$\, with $\Lie_{\ello}\, a = 0$,\, we have $V^a = \big(k \mathring{\bar{v}} + \mathring{\bar{s}}\big)\, \ello^a$ where $\mathring{\bar{s}} = \mathring{s} - k a$. On the other hand, if we were to replace the fiducial pair $(\qo_{ab},\, [\ello^a])$ we have used to describe the symmetry vector fields by another pair $(\qo^\prime_{ab}=\alpha^2 \qo_{ab},\, [\ello^{\prime\,a}] =  [\alpha^{-1}\ello^a])$, we would have $k^\prime = k$,\,\, $\vo^\prime \ello^{\prime \, a} =  \vo \ello^a$, \, and $\mathring{s}^\prime = \alpha\, \mathring{s}$. Thus, of the pair $(k, \mathring{s})$ labelling the vertical symmetry $V^a$,\, $\mathring{s}$ is a scalar field of conformal weight $1$, while the constant  $k$ is conformally invariant and is therefore denoted just by $k$ rather than $\mathring{k}$. (Recall that at null infinity $\scrip$, the supertranslations can be written as $s\, n^a$ and $s$ is also a function with conformal weight $1$.)

Note that the vertical vector fields constitute a sub-Lie algebra $\mathfrak{v}$ of $\g$, since the commutator of two vertical vector fields is again vertical:
\be \label{mathfrakv} [V_1,\,\, V_2]^a \, = \, (k_2\, \mathring{s}_1 - k_1 \mathring{s}_2)\, \ello^a\, . \ee 
While this sub-algebra is non-Abelian, it admits an infinite dimensional Abelian sub-Lie algebra $\mathfrak{s}$ consisting of vertical vector fields of the type $V^a =\mathring{s}\,\ello^a$ with $\Lie_{\ello}\,\mathring{s}=0$. Borrowing terminology from the BMS group, we will refer to these vector fields as \emph{supertranslations} and denote their Lie-algebra by $\mathfrak{s}$. Note that the Lie bracket (\ref{mathfrakv}) of any two vertical vector fields is a supertranslation. \smallskip

Let us next consider a general symmetry vector field $\xi^a$ and take its Lie bracket with a vertical vector field $V^a = f \ello^a\,$. Using (\ref{main}) we obtain:
\be \label{videal}[ \xi,\, V]^a \,=\, \big(\Lie_\xi \, f\, -\, (\phio+k)\, f \big)\, \ello^a\, .  \ee
The right side is a vertical vector field and hence in $\mathfrak{v}$. (In fact an  explicit calculation shows that the right side is a supertranslation. See Eq. (\ref{commutator}) below.) Thus $\mathfrak{v}$ is a Lie-ideal of $\g$.  

Let us take the quotient ${\mathfrak{g}} / {\mathfrak{v}}$. An element of the quotient is an equivalence class $[\xi^a]$ of symmetry vector fields $\xi^a$ where two vector fields are regarded as equivalent if they differ by a vertical symmetry. Therefore, each equivalence class has a unique projection $\ul\xi^a$ to the 2-sphere $\Houl$ and the correspondence $[\xi^a] \leftrightarrow \ul\xi^a$ is an isomorphism of Lie algebras. Now, the first of Eqs.  (\ref{main}) implies that $\ul\xi^a$ is a conformal Killing field on $(\Houl,\, \qoub{}_{ab})$. As is well known, the Lie algebra of conformal Killing fields on a round 2-sphere is isomorphic to the Lorentz Lie algebra $\mathfrak{l}$ (see, e.g. \cite{rpwr} or, for a summary, Remarks 1 and 2 of the Appendix). Hence the quotient Lie-algebra $\mathfrak{g}/\mathfrak{v}$ is isomorphic with $\mathfrak{l}$. \smallskip

To summarize, the Lie-algebra $\g$ is a semi-direct sum of the Lie-algebra $\mathfrak{v}$ of vertical vector fields and the Lorentz Lie algebra $\mathfrak{l}$. 
The finite diffeomorphisms generated of the vector fields $\xi^a$ constitute the symmetry group $\G$. It is the semi-direct product of the group $\mathfrak{V}$ of vertical diffeomorphisms generated by vector fields $V^a$ and the Lorentz group, in which $\mathfrak{V}$ is the normal subgroup: $\G = \Ver\ltimes  \Lor$. 

Finally, we chose to discuss the universal structure and NEH symmetries using an abstract 3-manifold $\Ho$. However, we could also work with 4-dimensional space-times $(M, g_{ab})$ admitting an NEH $\H$ as the inner boundary, and focus on the universal structure on $\H$. Let ${\rm Diff}_0 (M)$ be the subgroup of the diffeomorphism group of $M$ that preserves the universal structure. Let ${\rm Diff}_0^0 (M)$ be its subgroup consisting of diffeomorphisms that are identity on $\H$. Then $\G = {\rm Diff}_0 (M)/{\rm Diff}_0^0 (M) $.

Note that every Killing field of $(M, g_{ab})$ that is tangential to $\H$ leaves invariant the pair $(q_{ab}, [\ell^a])$ on $\H$ and therefore preserves the 3-parameter family of pairs $(\qo_{ab}, [\ello^a])$. Hence its restriction to $\H$ is an infinitesimal NEH symmetry. Thus space-time isometries that leave $\H$ invariant are guaranteed to be NEH symmetries, just as one would expect. \smallskip

\subsubsection{Salient properties of $\g$ and $\G$}
\label{s3.2.2}

Recall that the BMS group $\B$ is also a semi-direct product, $\B = \Super \ltimes \Lor$, of the supertranslation group $\Super$ with the Lorentz-group. Therefore, there is a close similarity between $\B$ and $\G$. However, as the explicit form of (\ref{vertical}) of vertical vector fields shows, the Lie-algebra $\mathfrak{v}$ is a 1-dimensional extension of the Lie-algebra $\mathfrak{s}$ of supertranslations. Interestingly, this simple extension endows $\g$ and $\G$ with a rich set of properties. We will now list the salient features that arise.

1. Let us ask if the Abelian sub-algebra $\mathfrak{s}$ of supertranslations is also a Lie-ideal. Given a supertranslation $S^a =\mathring{s}\ello^a$ (with $\Lie_{\ello}\mathring{s}=0$) and a general symmetry vector field $\xi^a$, we have:
\be \label{sideal} [ \xi, \, S]^a\, = \, \big(\Lie_{\xi}\,\mathring{s}- (\phio +c)\,\mathring{s}\big)\, \ello^a \, .\ee
It is easy to check that the coefficient $\big(\Lie_{\xi}\,\mathring{s}- (\phio +c)\,s \big)$ is Lie-dragged by $\ello^a$, whence the right side is a supertranslation. (Indeed,  this is just a special case of the commutator (\ref{videal})). Therefore the quotient $\q := \g/{\mathfrak{s}}$ is also a Lie algebra. Thus, $\g$ can also be written as a semi-direct sum of the Lie-algebra $\mathfrak{s}$ of supertranslations and the the Lie algebra $\q$.

As a vector space, each element of the quotient can be labelled by a pair $(k\vo\,\ello^a,\,\, \ul{\xi}^a )$ consisting of a vertical vector field $k\vo\,\ello^a$ on $\Ho$ and a conformal Killing field $\ul{\xi}^a$ on $\Houl$.   
\footnote{Note that if we were to replace the fiducial pair $(\qo_{ab},\, [\ello^a])$ we used to describe the symmetry vector fields so far by another pair $(\qo^\prime_{ab}=\alpha^2 \qo_{ab},\, [\ello^{\prime\,a}] =  [\alpha^{-1}\ello^a])$, we have: $k^\prime\vo^\prime\,\ello^{\prime\,a} = k\vo\,\ello^a$,\,\, $k^\prime = k$ and \,\, $\ul{\xi}^{\prime\,a} = \ul{\xi}^a$.}  
Thus, the quotient $\q$ is a seven dimensional Lie-algebra --\,``the Lorentz Lie algebra $\mathfrak{l}$, augmented with a one-dimensional vector field  $k \vo\, \ello^a$". Since this vector field can be written as $k \vo\, (\partial/\partial{\vo})$, we will refer to it as the `\emph{dilation vector field}' and denote it by $d^a$. Note that its commutator with any supertranslation $S^a =\mathring{s}\ello^a$ is given by: $[S, d]^a = k S^a$; action of $d^a$ just dilates the supertranslation by the constant $k$.

Let us explore the Lie algebra structure of $\q$. We begin by noting that since $\Lie_{\ello}\, \vo =1$, given any symmetry vector field $\xi^a$, one can show using  Eq.(\ref{main}) that $\Lie_\xi \vo = (c+\phio) \vo + \mathring{s}$ for some function $\mathring{s}$ satisfying $\Lie_{\ello}\mathring{s}=0$. Using this fact and Eq. (\ref{main}), it follows that the commutator has the form
\be [\xi,\, d ]^a \, = \, \mathring{\t{s}}\, \ello^a \, \qquad \hbox{\rm where $\mathring{\t{s}}$\,\, satisfies\,\, $\Lie_{\ello^a} \t{s} =0$}\, . \ee 
Thus, the commutator is a supertranslation, and hence in the kernel of the quotient. Therefore, the element $(k \vo\, \ello^a,\, 0)$ of $\q$ commutes with every other element of $\q$; it is in the center. Thus, $\q = \mathbb{R} \oplus \mathfrak{l}$;\,
it is a direct sum  of the (trivial) 1-dimensional Lie-algebra $\mathbb{R}$ with the Lorentz Lie algebra.\smallskip

2. Let us next consider the vector space $\mathfrak{b}$ of symmetry vector fields $b^a$ for which the constant $k$ in (\ref{main}) vanishes, so that:
\be \label{mainbms}\Lie_{b} q_{ab} = 2\phio_{(b)}\, q_{ab} \quad {\rm and} \quad \Lie_{b} \ello^a = -\phio_{(b)} \ello^a \ee
It is easy to verify that $\mathfrak{b}$ is closed under the Lie bracket operation. 
Indeed, Eqs. (\ref{mainbms}) are precisely the (direct analogs of) equations that define the generators of the BMS group (see, e.g., \cite{aa-yau}). Hence $\mathfrak{b}$ is naturally isomorphic with the BMS Lie algebra.

Let $\xi^a$ be a general symmetry vector field and consider its commutator\,\, $\eta^a:= [\xi, b]^a$ with a BMS vector field $b^a$. Then a simple calculation yields
\be \Lie_{\eta} \qo_{ab} = 2\phio_{(\eta)}\, \qo_{ab} \qquad {\rm and} \qquad 
     \Lie_{\eta} \ello^a = - \phio_{(\eta)}\,\ello^a \qquad {\rm where}\quad \phio_{(\eta)} = \Lie_{\xi} \phio_{(b)} - \Lie_{b} \phio_{(\xi)}\,  \ee
(where we have restored the suffix $\eta$ and $b$ on $\phio$ since we are dealing with two distinct symmetry vector fields). Hence $\eta^a$ belongs to $\mathfrak{b}$, whence $\mathfrak{b}$ is also a Lie-ideal in the symmetry Lie algebra $\g$. The quotient is the 1-dimensional vector space spanned by NEH symmetries of the type  $\xi^a = k \vo \,\ello^a$, i.e. by dilations. Thus, interestingly, $\g$ admits another semi-direct sum structure, \emph{now with the BMS Lie-algebra $\mathfrak{b}$ as the Lie-ideal.}\smallskip

3. As is well-known, the BMS Lie-algebra admits a 4-dimensional sub-algebra $\mathfrak{t}$ of translations. Therefore the NEH symmetry algebra $\g$ also admits a 4-dimensional Abelian subalgebra of translations. $\mathfrak{t}$\, is a sub-algebra of the supertranslation algebra $\mathfrak{s}$ where the (conformally weighted) coefficients $\mathring{s}(\theta,\varphi)$ satisfy: $\Do_a \Do_b\,\mathring{s}\,\propto \,\qo_{ab}$. Alternatively, they are linear combination of the first four spherical harmonics. (The conformal weight ensures that as we change $\qo_{ab}$ to $\qo^\prime_{ab}$, the 4-dimensional space is left invariant.) Let us denote the translation vector fields by $t^a$ and write them as $t^a = t\, \ello^a$. Since $[d, \, t \ello]^a = - k t\, \ello^a$, it follows that $\mathfrak{t}$ is a Lie ideal also of the NEH symmetry algebra $\g$. This is in striking contrast with the symmetry group associated with null surfaces discussed in \cite{cfp} which does not admit a translation subgroup. \smallskip

4. Let us provide explicit expressions of the symmetry vector fields $\xi^a$ by introducing a convenient chart. Fix a fiducial metric $\qoub{}_{ab}$ on $\Houl$ and  let $(\theta,\varphi)$ be the lifts to $\Ho$ of a set of spherical coordinates on $\Houl$ that are adapted to it. 
Fix any cross section $C_0$ of $\Ho$, and let $\vo$ be the affine parameter of a vector field $\ello^a$ in $[\ello^a]$ that vanishes on $C_0$. Then, every symmetry vector field $\xi^a$ is parametrized by \emph{a constant $k$ and functions $s,\, \chio,\, \phio$} that are lifts to $\Ho$ of functions on the 2-sphere $\Houl$:
\be \label{xi1} \xi^a = \big((k + \phio)\vo + s)\big) \ello^a \,+\, \epsilono^{ab} \Do_b \chio\, + \, \qo^{ab} \Do_b \phio\, . \ee
Here $\chio$ and $\phio$ satisfy (\ref{Y1m}), i.e., are linear combinations of the $\ell = 1$ spherical harmonics of $\qo_{ab}$, and  $\epsilono^{ab},\, \qo^{ab}$ are the contravariant area 2-form and metric on the $\vo= {\rm const}$ cross-sections, respectively. Given this explicit form, one can readily check that $\xi^a$ satisfies Eqs. (\ref{main}) that define infinitesimal horizon symmetries. 

In the decomposition provided by Eq. (\ref{xi1}), we can interpret various terms as follows: $d^a := k\vo\, \ello^a$ is a dilation (that is absent in the BMS Lie algebra $\mathfrak{b}$);\, $S^a:= \mathring{s} \ello^a$ a supertranslation;\, $R^a := \epsilono^{ab} \Do_b \chio$ a rotation; and $B^a := \qo^{ab} \Do_b \phio\, + \vo \phio\, \ello^a$\, a boost. Thus,
\be \xi^a\, =\, (d^a + S^a) + R^a + B^a\, \equiv\, V^a + R^a + B^a. \ee
However, this decomposition depends on the choice of the fiducial $\qo_{ab}$ and $\ello^a$ as well as the choice of the cross-section $C_0$. Nonetheless, it serves to bring out the `degrees of freedom' we have in the Lie algebra $\g$: a 1-dimensional freedom in $k$; a 3-dimensional freedom in each of $\phio$ and $\chio$;\, and an infinite dimensional freedom (worth a function on a 2-sphere) in the choice of $\mathring{s}$. The decomposition is also useful in exploring the Lie algebra structure of $\g$ explicitly, in particular the various semi-direct sums of Lie algebras involved. For this, it is simplest to use the expression of the commutator:
\be [\xi_1,\, \xi_2]^a = \xi_3^a   \ee
with 
\ba \label{commutator} d_3^a &=& 0, \quad  S_3^a = \big[\big((k_2+\phi_2) \mathring{s}_1 -(k_1+\phi_1) \mathring{s}_2\big) + (\Lie_{R_1} \mathring{s}_2 - \Lie_{R_2} \mathring{s}_1) + (\Lie_{B_1} \mathring{s}_2 - \Lie_{B_2} \mathring{s}_1)\big] \ello^a \nonumber\\
R^a_3 &=& \Lie_{R_1} R_2^a,  \qquad  B_3^a = \qo^{ab}\Do_a \phio_3 + \vo\,\phio_3 \ello^a \quad {\rm where}\quad  \phio_3 = (\Lie_{R_1 +B_1}\, \phio_2 -  \Lie_{R_2 +B_2}\, \phio_1)\, . \nonumber \\ \ea

5. Finally, we can summarize this rich structure by lifting the Lie algebra results of this section to groups. The symmetry group $\G$ is the subgroup of the diffeomorphism group of $\Ho$ that preserves the universal structure. It is infinite dimensional. It admits several normal subgroups. The first, $\Ver$ is generated by the vertical symmetry vector fields and is non-Abelian. The quotient $\G / \Ver$ is isomorphic to the Lorentz group $\Lor$. Thus, as we already noted in Sec.~\ref{s3.2.1}, $\G$ has a semi-direct product structure: $\G = \Ver\ltimes  \Lor$. 

The second normal subgroup is the Abelian group $\Super$ of supertranslation. The quotient $\G / \Super$ is a seven dimensional group $\Q$, the direct product $\Lor \times \mathbb{R}$ of the Lorentz group with the 1-dimensional Abelian group (generated by the dilation vector fields). Thus, the full symmetry group $\G$ is also realized as another semi-direct product: $\G = \Super \ltimes \Q$. A third normal subgroup is the BMS group $\B$, generated by vector fields $\xi^a$ with $k=0$. Now, the quotient $\G/\B$ is the 1-dimensional subgroup $\mathfrak{D}$ generated by dilations. Therefore $\G$ can also be written as $\G = \B \ltimes \mathfrak{D}$.

Recall that the BMS group admits a 4-dimensional normal subgroup $\T$ of translations, given by $\vo \to \vo\,+\, t(\theta, \varphi),\,\, \theta \to \theta,\,\, \varphi \to \varphi$, in the notation introduced above. Here $t(\theta, \varphi)$ are conformally weighted functions with weight $1$, defined by the property that $\Do_a \Do_b t \,\propto\, \qo_{ab}$ for \emph{any} unit, round, 2-sphere metric $\qo_{ab}$ in our collection. Thus, the NEH symmetry group $\G$ admits a 4-dimensional translation subgroup $\T$. Furthermore, $\T$ is a normal subgroup also of $\G$.

Thus, $\G$ has a rich and interesting structure. Different features of this structure becomes important in various applications. In particular, some play an  important role in the relation between NEHs and null infinity of asymptotically flat space-times discussed in the forthcoming papers, while others are needed to explore the physics and geometry of cosmological horizons \cite{aasb}.

\subsubsection{Comparisons}
\label{s3.2.3}

We will now compare the structure of $\G$ with other similar symmetry groups that have been discussed in the literature.
\smallskip

1. The difference between the structure of the NEH symmetry group $\G$ and the BMS group $\B$ at $\scri$ lies in the fact that $\G$ has an extra dimension encoded in the dilation group $\mathfrak{D}$. How did this arise? The defining equation of the NEH symmetries $\xi^a$ is (\ref{main}) while that of BMS symmetries is (\ref{mainbms}). The extra factor $- k\ello^a$ in (\ref{main}) arose because, associated with a round, unit metric $\qo_{ab}$ on $\Ho$ we only have an equivalence class $[\ello^a]$ of vector fields that are related by a rescaling by a positive constant, rather than a single vector field $\ell^a$.  As our discussion in Sec.~\ref{s2} showed, one cannot do better on an NEH. By contrast, at $\scri$, each $\qo_{ab}$ comes with a specific null normal.
\footnote{ Interestingly, extensions of the BMS group containing our symmetry group $\G$ have appeared in the literature, motivated by mathematical \cite{hawking2017} as well as Hamiltonian \cite{freidel2021} considerations at $\scrip$. They have also appeared in the context of black hole entropy calculations \cite{donnay2016}. Typically, these symmetry groups are significantly larger than our $\G$ and their structure does not have as close a relation to $\B$.}

Indeed, the 1-dimensional enlargement of the BMS Lie algebra $\mathfrak{b}$ through the presence of the dilation vector field $d^a$ is \emph{essential} from a space-time perspective. Consider a concrete NEH, say the future horizon in the Schwarzschild solution. Then, the restriction of the time translation Killing field to this NEH (which is also the event horizon in this case) is a dilation. Thus, had the extension from $\B$ to $\G$ not taken place, we would have found that an exact space-time symmetry is not an NEH symmetry!  

Consider the five dimensional group $\t{\T}$, generated by the dilation group $\mathfrak{D}$ and the 4-dimensional BMS translation subgroup $\T$. It may be appropriate to regard $\tilde{\T}$ as the `extended translation group' of NEHs. For, as we just noted, the restriction of the static Killing field to $\H$ in Schwarzschild space-time is the dilation, while in the extremal Reissner-Nordstrom space-time, it is an NEH time-translation. Similarly, on the cosmological horizon $E^+(i^-)$ of perturbed deSitter space-time (see the left panel of Fig.~\ref{positivecc}), restrictions of the three space-translations (on the Poincar\'e patch) of the background de Sitter metric are NEH space-translations that belong to $\T$ while the restriction of the time translation (of the static patch) is a dilation \cite{abk}. The five dimensional $\t{\T}$ is not Abelian, but is a semi-direct product of the Abelian, normal subgroup $\T$ with the 1-dimensional group $\mathfrak{D}$ of dilations:\, $\t{\T} = \T \ltimes \D$. In this semi-direct product, the action of the quotient on the normal subgroup is just to dilate the translations by a constant. \smallskip

2. NEHs were considered in a recent investigation of gravitational waves in presence of a positive cosmological constant as the relevant past null infinity $\scrim_{\rm Rel}$ to specify the `no-incoming radiation' condition in space-times representing isolated systems \cite{aasb}. In that analysis, the universal structure was chosen somewhat differently: It consisted of the 3-parameter family of unit, round metrics $\qo_{ab}$ as in the present case, but a single equivalence class of null-normals $[\ell^a]$, associated directly with the NEH, unaffected by the rescalings by $\psi$ that relate $q_{ab}$ to the $\qo_{ab}$. As we saw in Sec.~\ref{s2}, the strategy of rescaling $\ell^a$ used in this paper is natural from the viewpoint of multipoles (and could also have been used in \cite{aasb} without affecting the main conclusions).  Still, it is of interest to compare and contrast the symmetry group $\G_{\rm AB}$ obtained in \cite{aasb} with our group $\G$. Both are one dimensional extensions of the BMS group, provided by dilations $d^a$. 
The group $\mathfrak{V}$ generated by the vertical vector fields are isomorphic and the quotient $\G/\mathfrak{V}$ and $\G_{\rm AB}/\mathfrak{V}$ are isomorphic to the Lorentz group $\Lor$. \emph{However, the semi-direct product structure is different} --i.e. the action of the quotient group on the normal subgroups is different (for details, see Sec.~ IV B of \cite{aasb}). In particular, the BMS group $\B$ is not a subgroup of $\G_{\rm AB}$. 

Since the difference can be traced back to the universal structures chosen in the two sets of analyses, it is natural to ask whether we could not add the equivalence class $[\ell^a]$ to the universal structure considered in Sec. \ref{s3.1}. After all, we should include all universally available fields in specifying the universal structure. However, this is not possible. On the abstract manifold $\Ho$, following \cite{aasb}, one can fix an equivalence class $[\ell^a]$ of vertical vector fields if one wishes: Since any two $\ell^a$ in the class differ just by a constant rescaling, relation between them can be specified universally. Similarly in our analysis we can specify the equivalence classes $[\ello^a]$'s since, as emphasized in Sec. \ref{s3.1}, the relation between any two $[\ello^a]$'s is again universal (i.e. does not refer to the physical metric $q_{ab}$ that varies from one NEH to another). However, the relation between $[\ell^a]$ and the $[\ello^a]$'s involves the conformal factor $\psi$ that relates $\qo_{ab}$ to $q_{ab}$ that varies from one NEH to another. Therefore, we cannot include both $[\ell^a]$ and $[\ello^a]$'s in the universal structure. One has to make a choice. We believe that the choice made in Sec.~\ref{s3.1} is better suited for applications because: (i) it is natural for defining multipoles; \, (ii) the direct relation to the BMS group is likely to be important in both classical and quantum considerations;\, and (iii) the main results of the analysis \cite{aasb} of gravitational waves in presence of a positive cosmological constant will go through with this choice. 

3. There is another setting in which the symmetry group associated with an inner boundary was discussed recently \cite{cfp}. In that analysis, the inner boundary $\N$ is required only to be null; it does not have to be an NEH. Therefore, the universal structure on $\N$ is different from that on our $\H$, and hence the symmetry group $\G_{\rm CFP}$ is also quite different from our $\G$\, (or, $\G_{\rm AB}$). Because this analysis allows boundaries that are much more general than our NEHs, $\G_{\rm CFP}$ is significantly larger than $\G$. First, in effect, our constant $k$ is replaced by a function on a 2-sphere; thus the sub-algebra $\mathfrak{V}_{\rm CFP}$ of vertical symmetries is ``worth \emph{two functions} on a 2-sphere''. Furthermore the quotient $\G_{\rm CFP} /\mathfrak{V}_{\rm CFP}$ is also infinite dimensional: it is the group ${\rm Diff}\, (\mathbb{S}^2)$ rather than the Lorentz group $\Lor$. As a result, apart from the fact that they both admit a semi-direct product structure, there is no direct relation between $\G_{\rm CFP}$ and the BMS group $\B$. In particular, $\G_{\rm CFP}$ does not admit a natural translation subgroup, while $\G$ does. In spite of these conceptual differences in kinematics, as we discuss in the companion paper \cite{akkl2}, results from the phase space analysis in \cite{cfp} can be taken over in our framework: Since that framework encompass all space-times with null boundaries, the results are in particular applicable to space-times in which the null boundary is an NEH, as well as the first order perturbations on such space-times.

\section{Discussion}
\label{s4}

Let us begin with a summary of the main results. In Sec. \ref{s2} we introduced multipole moments $I_{\ell, m}$ and $L_{\ell,m}$ that characterize the geometry of any given NEH $\H$. The $I_{\ell, m}$ capture the distortions in the intrinsic horizon geometry, encoded in the scalar curvature of its metric (which equals ${\rm Re} \Psi_2$ on $\H$), while $L_{\ell,m}$ capture the angular momentum structure of the horizon, encoded in the `rotation 1-form' $\omega_a$ (that serves as a potential for ${\rm Im} \Psi_2$). To define these multipoles, one needs spherical harmonics. In the previous work, these were obtained by first restricting oneself to axisymmetric NEH geometries and then using the fact that every axisymmetric metric $\qub{}_{ab}$ on $\mathbb{S}^2$  naturally defines a canonical round 2-sphere metric $\qoub{}_{ab}$ that shares the rotational Killing vector of $\qub{}_{ab}$ \cite{aepv}. We removed the restriction to axisymmetric geometries by adopting a new approach based on the following observation: Given any $\qub{}_{ab}$ on $\mathbb{S}^2$, there  exists a unique 3-parameter family of unit, round 2-sphere metrics $\qoub{}_{ab}$, conformally related to the given $q_{ab}$. Each of these $\qoub{}_{ab}$ defines a set of spherical harmonics which can be used to extract the geometrical multipole moments from the given NEH geometry. We then showed that, given just the set of numbers $I_{\ell, m}$ and $L_{\ell, m}$ extracted from an NEH geometry, one can systematically reconstruct that geometry. Furthermore, given any set of these numbers with an appropriate fall-off for large $\ell$ for the spherical harmonic expansion to converge (see Eq. (\ref{EB})), the construction provides an NEH geometry; additional weak but rather awkward conditions on multipoles that were necessary to ensure axisymmetry in the previous work are no longer needed. Also, in that work only the $m=0$ multipoles were non vanishing. Now, the set of multipoles is richer: NEHs can have non vanishing multipoles for all $\ell, m$. 

Note that to begin with we have a 3-parameter family of multipoles, rather than a single set, each defined by a round metric $\qoub{}_{ab}$. As explained in the Appendix, the round metrics are related by the action of the Lorentz group and multipoles transform in a well-defined manner as one changes the round metric. The non-uniqueness of multipoles is not surprising: one knows already from multipoles characterizing static solutions of Einstein's equations that there can be distinct invariant characterizations of a geometry \cite{geroch,hansen}. Nonetheless, in practice it is much more convenient to fix the `gauge freedom' in the choice of the round metric. We provided one prescription to select a \emph{canonical} round metric $\qout{}_{ab}$, discussed in detail in the Appendix (which also has the property that if the metric $q_{ab}$ on a given NEH admits an axisymmetric Killing field $R^a$, then $R^a$ is also a Killing field of  $\qout{}_{ab}$). The resulting set of multipoles provide an invariant characterization of the horizon geometry that can be used, e.g., to compare the horizon structures that result in numerical simulations 
that use different coordinates systems, tetrads and gauge fixing procedures (see, in particular \cite{schnetter_2006,Gupta_2018,pookkolb2020horizons,prasad2021tidal}). 

The discussion that led to multipoles brought out the universal structure of  NEHs, i.e. structure that is shared by all NEHs and can therefore be specified on an abstract 3-manifold $\Ho = \mathbb{S}^2 \times \mathbb{R}$, without reference to fields on any specific NEH. It consists of a 3-parameter family of  pairs $(\qo_{ab},\, [\ello^a])$ of unit round metrics, and equivalence classes of vector fields that rule $\Ho$, where $\ello^{a} \sim  c \ello^a$ with $c$ is a positive constant. Any two pairs in this collection are related by a conformal factor: $\qo^{\prime}_{ab} = \alpha^2 \qo_{ab}$ and $[\ello^{\prime\, a}] = [\alpha^{-1} \ello^{\prime\, a}]$ where the 3-parameter freedom in the conformal factor $\alpha$ can be made manifest, as in Eq. (\ref{alpha}). The group $\G$ that preserves this structure is the NEH symmetry group. We explored its structure in detail in Sec. \ref{s3.2}. Interestingly, $\G$ is a one dimensional extension of the BMS group $\B$ at $\scri$: $\B$ is a normal subgroup of $\G$ and the quotient $\G/\B$ is isomorphic to the additive group $\mathfrak{R}$ of reals. One can also understand this difference as a 1-dimensional extension of the  supertranslation subgroup $\Super$ of the BMS group $\B$ by a `dilation' whose action simply rescales the supertranslations. In terms of Lie-algebras, the `vertical' symmetry vector fields $V^a$ are now given by $V^a = d^a + S^a \equiv k \vo\, \ello^a + S^a$, where $k$ is the constant that characterizes the dilation and $S^a = \mathring{s}(\theta,\varphi) \ello^a$ is a supertranslation as on $\B$. (If one uses the conformal frame $\big(\qo^{\prime}_{ab},\, [\ello^{\prime\, a}]\big)$ instead, the labels $(k, \mathring{s})$ of $V^a$ transform as $(k^\prime = k,\, \mathring{s}^\prime = \alpha\, \mathring{s}$; thus as in the BMS group, $\mathring{s}$ has conformal weight 1, while the label $k$ accompanying the dilation is conformally invariant.) This 1-dimensional enlargement of the supertranslation group by the dilation vector field is a direct consequence of the fact that we only have an equivalence class $[\ello^a]$ of normals in any conformal frame at NEHs rather than a specific null normal as at $\scri$. This in turn is forced on us by the fact that any concrete NEH is endowed only with an equivalence class $[\ell^a]$ of preferred null normals (selected by the condition that the `rotation' 1-form $\omega_a$ be divergence-free) rather than a specific null normal $\ell^a$. While surprising at first, this enlargement cannot be avoided: If the NEH is a (non-extremal) Killing horizon --say, as in the Schwarzschild space-time-- then the restriction of the Killing vector to the NEH is a dilation! Thus, had dilation been absent in the symmetry group $\G$, one would have have been led to the conclusion that isometries of a space-time need not belong to the group $\G$ of NEH symmetries even when they leave the NEH invariant. This does not occur: Just as every space-time Killing field defines an infinitesimal symmetry at $\scri$ in the asymptotically flat case, every space-time Killing vector (that is tangential to an NEH) defines an infinitesimal NEH symmetry.

In the companion paper \cite{akkl2}, we will use the infinitesimal generators $\xi^a$ of the symmetry group $\G$ to define charges $Q_\xi [C]$ associated with cross-sections $C$ and the corresponding fluxes $F_\xi [\N_{1,2}]$ across the region $\N_{1,2}$ bounded by two cross-sections $C_1$ and $C_2$ of a perturbed NEH. A general covariant phase space framework for manifolds with internal null boundaries $\N$ was developed in \cite{cfp}. It is mathematically attractive because it places no restrictions on the boundary other than requiring it to be smooth and null. On the other hand, this very generality allows situations in which charges and fluxes have physically undesirable properties. For example, the framework allows one to take $(M, g_{ab})$ to be the exterior of the null cone of a point in \emph{Minkowski space} (or a part thereof), with the null cone (or part thereof) serving as the boundary $\N$. But then the charge \emph{and the flux}  associated with the dilation vector field (which is also a symmetry in the Lie-algebra $\g_{\rm{CFP}}$ of \cite{cfp}) are non-zero. Given that space-time under consideration is \emph{flat}, it is difficult to associate physical significance to these charges and, especially, fluxes. More generally, the fluxes associated with symmetry vector fields which are time-like near $\N$, and future directed and null everywhere on $\N$, can be negative. We will find that these undesirable features disappear if one restricts $\N$ to be an NEH or a perturbed NEH. As noted in Sec.~\ref{s1}, perturbed NEHs are meant to approximate dynamical horizons (DHs) in regimes in which fields are slowly varying, i.e. when the flux of matter and gravitational radiation falling into the horizon is weak. To encompass fully dynamical situations, it seems necessary to abandon null boundaries and use instead dynamical horizons (DHs) as interior boundaries.%
\footnote{Similarly, for cosmological horizons in the positive $\Lambda$ case, depicted in the left panel of Fig.~\ref{positivecc},\, $E^{-}(i^{+})$ would be a perturbed NEH --and serve as ``local $\scrip$" for the compact binary \cite{aasb}-- only if the shear is sufficiently weak, as, e.g, in \cite{kl2020}, for the past null geodesics from $i^{+}$ to reach $E^+(i^-)$ without encountering caustics. If the emitted radiation is strong then, presumably, one would have to abandon ``local $\scrip$" and work with only the global, space-like $\scrip$ shown in the figure.} 

But what about event horizons (EHs), then? A covariant phase-space description with EHs as inner boundaries faces two obstacles. The first is a mathematical difficulty already noted in \cite{cfp}: the Raychaudhuri equation implies that generically EHs will not be smooth boundaries because of conjugate points. This will likely make the task of defining charges and fluxes quite difficult. The second problem stems from physics. As the Vaidya solution illustrates, EHs can form and grow in a flat region of space-time (see, e.g., \cite{ak-dh,akrev}). In this region the EH is smooth. This suggests that it would be difficult to attribute physical significance to  charges and fluxes, even if one were to restrict oneself to portions of EH that are smooth. If, instead,  one were to extend the Hamiltonian framework with a DH as the internal boundary, both difficulties would disappear in the full non-perturbative regime. The issue of non-smoothness will not arise because DHs are smooth and have no problems associated with geodesic crossing. The physical difficulties would also disappear, since the growth of the area of 2-sphere cross-sections of a DH is directly related to the flux of energy falling across it \cite{ak-dh}. Thus, an extension of the present framework to space-times with DHs as internal boundaries appears to be perhaps the most promising avenue to extend the rich interplay between geometry and physics to full, non-linear situations, e.g., in black hole mergers.

\section*{Acknowledgment}
                We thank Alexandre Le Tiec for discussions and especially for drawing our attention to the
typos in sections 2.2 and 2.3.1. This work was supported by the NSF grant PHY-1806356, the Eberly Chair funds of Penn State, and the Mebus Fellowship to NK. MK was financed from budgetary funds for science in 2018-2022 as a research project under the program "Diamentowy Grant". MK and JL were supported by Project OPUS 2017/27/B/ST2/02806 of Polish National Science
Centre.

\begin{appendix}

\section{Selecting a canonical round metric}
\label{a1}

Recall from Sec.~\ref{s2.2} that every NEH admits a 3-parameter set $\S$ of round metrics $\qo_{ab}$, conformal to the physical metric $q_{ab}$.%
\footnote{For notational simplicity, in this Appendix we will drop the distinction between fields on the 2-sphere $\Hul$ and their lifts to the NEH $\H$. The integrals below can either be over $\Hul$, or, over any cross-section of $\H$.}
Each $\qo_{ab}$ provides the basis functions $\Y_{\ell,m}$ that can be used as `weights' to define the geometrical multipoles $I_{\ell,m},\, L_{\ell,m}$. To get a `canonical' set of multipoles, then, we need to choose a canonical round metric $\qout{}_{ab}$ in $\S$. As explained in Sec.~\ref{s2.3}, the idea is to select this $\qout{}_{ab}$  by requiring that the `area dipole moment' should vanish (see Eq.~\eqref{cm}). The purpose of this appendix is to show that the required $\qout{}_{ab}$ exists for any $q_{ab}$ on $\H$ and, furthermore, it is unique.

Let us begin with the issue of existence of $\qout{}_{ab}$. Fix a fiducial unit, round metric $\qo_{ab}$ on the 2-sphere $\Hul$ and regard this 2-sphere as being embedded in $\mathbb{R}^3$, so that points on $\Hul$ are labeled by coordinates $r^i = (x,y,z) =(\sin\theta\cos\varphi,\,\sin\theta\sin\varphi,\, \cos\theta)$. It turns out to be more convenient to use the real triplet of functions\, $r^i$\, on $\Hul$ in place of the $\Y_{1,m}$ defined by $\qo_{ab}$. In terms of these $r^i$, the condition \eqref{cm} to select $\qout{}_{ab}$ can be rewritten as
    \begin{equation}
         \label{cm_cartesian}
         \oint r^i\, \rmd^2 V\,\, =\,\, \oint {r^i}\, \psiut^{-2} \, 
         \dvout\, =\, 0 \, ,
    \end{equation}  
where $\rmd^2 V$ and and $\dvout$ are the area elements defined by the physical metric $q_{ab}$ and $\qout_{ab}$ respectively. The dipole moment $d^{\,i} := \oint r^i\, \rmd^2 V$\, provides a map $\M$ from the set $\S$ of round metrics to a subset $\D$ of $\mathbb{R}^3:\,\, \M (\qo) = d^{\,i}$. Since the fiducial metric $\qo_{ab}$ is arbitrary, generically $\M (\qo)$ will not vanish. We seek a $\qout{}_{ab}$ in $\S$ such that $\M (\qout) = 0$.

Now, as we indicated in Sec.~\ref{s2.2}, there is a 1-1 correspondence between the 3-dimensional space of round metrics $\qo_{ab}$ and the hyperboloid of unit time-like vectors in Minkowski space (see Eq. (\ref{alpha}) and also Remark 2 below). Moreover the conformal isometry group of each of these metrics is isomorphic with the Lorentz group (see, e.g., \cite{rpwr}). The fiducial $\qo_{ab}$ is invariant under an $SO(3)$ subgroup. Therefore, to find the desired mapping from the fiducial $\qo_{ab}$ to  $\qout{}_{ab}$ we can focus on the (complementary) 3-dimensional subset that corresponds to boosts. This subset is labeled by rapidity $\vec{\eta} \in \mathbb{R}^3$. Applying these transformations to the fiducial $\qo_{ab}$ we get a three parameter family of metrics $\qo_{ab}(\vec{\eta})$. Each of these metrics has the associated triplet $r^i_{\vec{\eta}}$, related to the $r^i$ determined by $\qo_{ab}$ via
    \begin{equation}
        r^i_{\vec\eta}\, =\, \frac{\Lambda^i{}_0 + \Lambda^i{}_j\, r^j
        }{\Lambda^0{}_0 + \Lambda^0{}_j\, r^j} \, ,
    \end{equation}
where $\Lambda^\mu_\nu$ is given by the Lorentz transformation with rapidity $\vec{\eta}$, and indices $i,j,$ run over $1,2,3$ (see Remark 1). In particular if we take $\vec{\eta} = \eta \hat{z}$ we get the following transformation:
    \begin{align}
        z_\eta& = \frac{z\cosh\eta +              
        \sinh\eta}{\cosh\eta+z\sinh\eta}\label{eqn:Lorentz_1}\\
        x_\eta& = \frac{x}{\cosh\eta+z\sinh\eta }\\
        y_\eta& = \frac{y}{\cosh\eta+z\sinh\eta }\label{eqn:Lorentz_3}
    \end{align}
Consequently, in the limit $\eta\to\infty$, we have $x_\eta, y_\eta \to 0$, and $z \to 1$, pointwise. Therefore we can take the limit under the integral in $\oint{r^i}_\eta \,\rmd^2 V$ to obtain  $\lim_{\eta\to\infty}\, \oint {r^i}_\eta\, \rmd^2 V = A_{\Hul}\, \hat{z}^i \equiv A_{\Hul}\, \delta^i{}_3$,\, where $A_{\Hul} = \oint \rmd^2 V$ is the area of $\Hul$ in the physical metric $q_{ab}$. Generalizing the boost to arbitrary directions in the $\vec\eta$ space, we have that 
   \begin{equation}
       \label{eq:lim_eta}
      \lim_{|\vec{\eta}|\to\infty}\, d^{\,i} (\qo_{\vec{\eta}})\,\, \equiv \,\,
        \lim_{|\vec{\eta}|\to\infty} \oint {r^i}_{\vec\eta}\, \rmd^2 V\,\, =\,\, 
       A_{\Hul}\,\hat{\eta}^i\, .   \end{equation}
Thus the map $\vec\eta \to  d^{\,i}(\qo_{\vec{\eta}})$ from the space of rapidities $\vec{\eta}$ to the set $\D$ of area dipole moments $d^{\,i}$ is such that the limiting 2-sphere of infinite radius in the space of rapidities is sent to the 2-sphere of radius $A_{\Hul}$ in $\D$. Note that this sphere in $\D$ encloses the point $d^{\,i} = 0$. This fact will lead us to a proof of the existence of a $\qout_{ab}$ with $d^{\,i}(\qout)=0$, using a higher dimensional analog of the intermediate value theorem based on homotopy. The idea is simple: If one continuously deforms a topological sphere that encloses a point $p$ into another that doesn't enclose it, $p$ must lie on an intermediate 2-sphere.  

Let us return to the fiducial $\qo_{ab}$, which corresponds to $\vec{\eta} =0$. If $d^{\,i}(\qo) =0$, then $\qo_{ab}$ itself is the required $\qout{}_{ab}$. If $d^{\,i}(\qo) \not=0$, then we will not reach the point $d^{\,i}=0$ by performing boosts with sufficiently small rapidity since the map $\vec\eta \to d^{\,i}(\qo_{\vec\eta})$ is continuous. Thus, there exists a 2-sphere of radius $\epsilon$ in the space of rapidities such that the 2-sphere in $\D$ consisting of points $d^{\,i}(\qo_{\vec\eta})$ with $|\eta| = \epsilon$ \emph{does not contain} the point $d^{\, i} =0$ in its interior. As we increase $|\eta|$ from $\eta =\epsilon$, in the limit $\eta\to \infty$ we obtain a 2-sphere in $\D$ that \emph{does contain} the point $d^{\,i} =0$ in its interior. Therefore, as $|\vec\eta|$ expands, the image in $\D$ of the 2-sphere $|\vec{\eta}| = {\rm const}$ must encounter the point $d^{\,i} =0$. That is, there exists an intermediate $\vec{\eta_\circ}$ such that $d^{\,i}(\qo(\vec{\eta}_{\circ})) = 0$. Our desired metric $\qout{}_{ab}$ is then this $\qo_{ab}({\vec{\eta}_\circ})$.

This proof of existence is very general in the sense that it continues to hold if the area dipole moment $d^{\,i}= \oint r^i  \rmd^2 V$ is replaced by another one, say $\tilde{d}^{\,i} = \oint r^i f\, \rmd^2 V$, so long as the corresponding `monopole' $\oint f \rmd^2 V$ does not vanish. This is precisely what one would expect from multipoles in Newtonian gravity or electrodynamics. For example, we can use the mass dipole moment for which $A_{\Hul}$ would be replaced by the mass monopole. (However, we cannot use the procedure to set the angular momentum dipole to zero because the angular momentum monopole moment vanishes.) \smallskip  

Let us now come to the issue of uniqueness of $\qout{}_{ab}$.  Let there be two round metrics satisfying Eq.~\eqref{cm_cartesian}. Then there is a conformal transformation given by some $\vec{\eta}_1$ that relates them. Without loss of generality we can take  $\vec{\eta}_1$ to be along the z-axis, so that  $\vec{\eta}_1 = \eta_1 \hat{z}$. We will now show that $\eta_1$ must be $0$. Let $\eta_t$ = $t\eta_1$ for $t\in[0,1]$. If $\eta_1>0$, then using (\ref{eqn:Lorentz_1}) we have
   \begin{equation}
       \frac{d}{dt}\oint z_{\eta_t}\, \rmd V\,\, =\,\, \eta_1\oint 
       (1-z_{\eta_t}^2) 
       \, \rmd V > 0
   \end{equation}
where, in the last step we have used $1- z_{\eta_t}^2 = x_{\eta_t}^2+y_{\eta_t}^2$. Therefore if $\eta_1 \not=0$, the integral is an increasing function of $t$ and cannot return to $0$ at $t=1$.  Thus, for any metric $q_{ab}$ on a sphere there is a 
unique, unit round metric $\qo_{ab}$ conformal to it such that its $\ell=1$ spherical harmonics satisfy Eq.~\eqref{cm}.
   
Note that this proof of uniqueness makes a strong use of positivity of the area 2-form, i.e., the fact that the integral of $\epsilon_{ab}$ over \emph{any} subregion of $\Hul$ is positive. For the mass dipole moment $I^i \sim \oint \mathcal{R}\,\, \rmd^2 V$, the proof would go through only if $\mathcal{R}\, \ge\, 0$ everywhere on $\Hul$. In the Kerr family this requirement is satisfied  if and only if $a \le (\sqrt{3}/2) M$ (i.e., if and only if the horizon radius $R_{\rm Hor}$ is between $2M$ and $3M/2$). However, we do not know if there is an alternate argument --perhaps with a natural additional requirement-- that would ensure uniqueness even in the case when the positivity fails.\\

\emph{Remarks:}\smallskip

1. The isomorphism between the Lorentz group $\mathfrak{L}$ and the group $\mathfrak{C}$ of conformal isometries of a unit, round, 2-sphere metric $\qo_{ab}$ can be geometrically visualized as follows \cite{rpwr}. Consider the null cone of the origin in Minkowski space and the 2-sphere cross-section $C$ defined by its intersection with the $t=1$ plane. Under a Lorentz transformation $\Lambda$ the null cone is left invariant but the cross-section is generically mapped to another cross-section $C^\prime$ (the intersection of the $t^\prime = 1$ plane with the null cone, where $t^\prime$ is the image of $t$ under $\Lambda$). The null rays generating the null cone provide a natural diffeomorphism between $C$ and $C^\prime$. If one pulls-back the metric on $C^\prime$ to $C$, we obtain a metric $\qo^\prime_{ab}$ on $C$ that is conformally related to $\qo_{ab}$. Since the Lorentz transformation is an isometry in Minkowski space, the scalar curvature of the 2-metrics on $C$ and $C^\prime$ is the same and constant (namely, $2$). Hence it follows that $\qo^\prime_{ab}$ is also a unit, 2-sphere metric. Therefore, there is a unique conformal isometry on $C$ that sends $\qo_{ab}$ to $\qo_{ab}^\prime$. This is the element of $\mathfrak{C}$ that corresponds to $\Lambda \in \mathfrak{L}$. Note that the $SO(3)$ subgroup of $\mathfrak{L}$ that leaves the vector $t^a \equiv (1,0,0,0)$ invariant maps $C$ to itself and therefore induces proper isometries of $\qo_{ab}$. The `genuine' conformal transformations in $\mathfrak{C}$ correspond to `pure' boosts in the rest frame in Minkowski space, defined by $t^a$.\smallskip

2. Next, let us elaborate on the relation between the space $\S$ of conformally related, unit 2-sphere metrics $\qo_{ab}$ and the hyperboloid $\mathcal{H}$ spanned by the unit time-like vectors in Minkowski space.  Any two metrics on $\S$ are related by a conformal isometry, and any two points on $\mathcal{H}$ by a Lorentz transformation. As we saw in Remark 1, the two groups $\mathfrak{C}$ and $\mathfrak{L}$ are isomorphic. The little group of any point $\qo_{ab}$ of $\S$ is an $SO(3)$ subgroup of $\mathfrak{C}$ and the little group of any point $p$ is an $SO(3)$ subgroup of $\mathfrak{L}$. Fix a fiducial metric $\qo_{ab}$ in $S$, and a point $p$ on $\mathcal{H}$, say with coordinates $(1,0,0,0)$ in Minkowski space. We can set up a 1-1, onto map $\pi$ from $\S$ to $\mathcal{H}$ as follows. Let us first set $\pi (\qo) = p$. Next, note that the isometry group of $\qo_{ab}$ is the $SO(3)$ subgroup of $\mathfrak{C}$, and the rotation group in the $x,y,z$ 3-space leaves $p$ invariant. Consider the 3-parameter families of `genuine' conformal transformations in $\mathfrak{C}$ and boosts in $\mathfrak{L}$ that are complementary to these $SO(3)$ subgroups. The action of these families extends the map $\pi$ providing us a 1-1 onto correspondence between $\S$ and $\mathcal{H}$.\smallskip

3. How natural is the requirement that the `area dipole' vanish? One test of naturalness comes from isometries: If $K^a$ is a Killing vector of the physical metric $q_{ab}$, is it also a Killing vector of the `canonical' metric $\qout{}_{ab}$ selected by our procedure?

Suppose the physical metric $q_{ab}$ itself is spherically symmetric. Then, if $R_{\Hul}$ is the horizon radius (w.r.t. $q_{ab}$),\,  $\qo_{ab} := (1/R^2_{\Hul})\,q_{ab} \in \S$, the space of unit, round 2-sphere metrics conformally related to $q_{ab}$. But it is the `canonical' round metric? The answer is in the affirmative since this the area dipole moment moment $d^{\,i}$ of $\qo_{ab}$ vanishes trivially. 

Next, let us consider an axisymmetric NEH so that $q_{ab}$ admits a rotational Killing field $R^a$. Since the diffeomorphism generated by $R^a$ leaves $q_{ab}$ invariant, it follows that it must preserve the 3-parameter family $\S$ of unit, round 2-sphere metrics. Therefore it is a conformal isometry of the preferred $\qout{}_{ab}$. But is it an isometry? We will now show that the answer is again in the affirmative. The argument is in three steps. First, consider the 1-parameter family of diffeomorphism $\varphi_t$ generated by $R^a$ and denote by $\qo_{ab}(t)$ the image of our preferred $\qout{}_{ab}$. Then each eigenspace of $\Dout^2$ with eigenvalue $\ell(\ell+1)$ is mapped to the eigenspace of $\Do^2(t)$ with the same eigenvalue, and this is an an isomorphism between the two subspaces. Consequently every $\Y_{1,m}(t)$ is the image of a linear combination $\sum\beta_m (t) \Yut{}_{1,m}$. Second, since the volume element $\rmd^2 V$ of $q_{ab}$ is preserved by $\varphi_t$, we have $\oint \Y_{1, m}(t)\, \rmd^2V = \oint \big(\sum_m \beta_m(t) \Yut_{\ell,m}\big)\,\rmd^2V=0 $\, where in the last step we have used the fact that the area dipole moment vanishes for $\qout{}_{ab}$. Since this equality holds for $m=0,\, \pm1$;  the area dipole moment w.r.t. $\qo_{ab}(t)$ also vanishes. Finally, recall that there is a unique unit, round metric $\qout_{ab}$ with this property. Hence $\qo_{ab}(t) = \qout{}_{ab}$; i.e., $\varphi_t$ is a Killing vector of $\qout{}_{ab}$.\smallskip

4. In the Kerr family, the conformal factor $\psiut$ relating the physical NEH metric $q_{ab}$ to the `canonical' round metric $\qout{}_{ab} := \psiut^2 q_{ab}$ is given, in the Boyer-Lindquist coordinates, by
\ba \psiut &=& \f{\sqrt{(r^2+a^2\cos^2\theta)(1- Z^2)}}{2Mr\sin\theta}\, ,\qquad  
{\rm where} \qquad \nonumber\\
Z &=& \tanh \Big(\f{2M \big(a\tan^{-1}(\f{a\,\cos\theta}{r}) + r \tanh^{-1}\, (\cos\theta)\big)}{a^2+r^2}\Big) \ea
Note that in the Schwarzschild space-time, $Z= \cos\theta$.
\smallskip

5. After this work was completed we became aware of a paper \cite{Gittel_2013} which provides a different proof of the main result of this Appendix. Since the motivation of that work was quite different, the proof is also quite different, with emphasis on analysis involving H\"older spaces, rather than on the direct geometric interplay between the action of the Lorentz group on a null cone and the conformal isometries of its 2-sphere cross-sections.

\end{appendix}

\bibliographystyle{JHEP}
\bibliography{bibl1}

\end{document}